\documentstyle[11pt,a4,cite,epsf,axodraw]{article}
\newcommand{\be}{\begin{equation}}
\newcommand{\ee}{\end{equation}}
\newcommand{\bea}{\begin{eqnarray}}
\newcommand{\eea}{\end{eqnarray}}
\newcommand{\noi}{\noindent}
\newcommand{\nn}{\nonumber}
\newcommand{\ra}{\rightarrow}
\newcommand{\cO}{{\cal O}}
 
\newcommand{\eq}[1]{Eq.~(\ref{#1})}

\newcommand{\NPB}[3]{Nucl.\ Phys.\ {\bf B{#1}} (19{#2}) {#3}}
\newcommand{\PRD}[3]{Phys.\ Rev.\ {\bf D{#1}} (19{#2}) {#3}}
\newcommand{\PLB}[3]{Phys.\ Lett.\ {\bf B{#1}} (19{#2}) {#3}}
\newcommand{\PRL}[3]{Phys.\ Rev.\ Lett.\ {\bf {#1}} (19{#2}) {#3}}

\newcommand{\AP}[3]{Ann.\ Phys.\ {\bf {#1}} (19{#2}) {#3}}
\newcommand{\ZPC}[3]{Z.\ Phys.\ {\bf C{#1}} (19{#2}) {#3}}
 
\newcommand{\sw}{s_{\rm w}}
\newcommand{\cw}{c_{\rm w}}
\newcommand{\swR}{s_{R,{\rm w}}}
\newcommand{\cwR}{c_{R,{\rm w}}}
\newcommand{\swe}{s_{\rm w,eff}}

\newcommand{\swb}{\bar{s}_{\rm w}}
\newcommand{\cwb}{\bar{c}_{\rm w}}
 
\newcommand{\alphaR}{\alpha_{R}}
\newcommand{\gR}{g_{R}}
\newcommand{\gpR}{g_{R}'}
\newcommand{\alphab}{\bar{\alpha}}
\newcommand{\gb}{\bar{g}}
\newcommand{\gpb}{\bar{g}'}
 
\newcommand{\ks}{{\mbox k \!\!\! /}}
\newcommand{\ps}{{\mbox p \!\!\! /}}
 
\newcommand{\im}{{\rm Im\,}}
\newcommand{\re}{{\rm Re\,}}

\def\theequation{\arabic{section}.\arabic{equation}}
 
\begin{document}
 
\begin{titlepage}
%\begin{flushleft} {\Large\sl Revised version}
%\end{flushleft}
\begin{flushright} CPT-96/P.3408\\ November 1996 \\ (revised May 1997)
\end{flushright}
\vspace*{2cm}
 
\begin{center}
 
{\LARGE {\bf Electroweak Effective Charges}}

\vspace{5pt}

{\LARGE {\bf and their relation to}}
 
\vspace{5pt}

{\LARGE {\bf Physical Cross Sections}}
\\[1.5cm]
{\large 
{\bf J. Papavassiliou}\footnote{Present address:
Dept.\ of Physics and Astronomy, 
%Schuster Laboratory,
University of Manchester, Manchester M13 9PL, UK.},
{\bf E. de Rafael} and 
{\bf N.J. Watson}\footnote{Present address:
%Division de Physique Th\'eorique, 
Institut de Physique Nucl\'eaire, 
Universit\'e de Paris-Sud, F-91406 Orsay Cedex, France.}
}\\[1cm]
 
\noindent
Centre  de Physique Th\'eorique, \\
CNRS-Luminy, Case 907, \\ 
F-13288 Marseille Cedex 9, France.

\end{center}
 
\vspace*{1.5cm}
\begin{abstract}

In quantum electrodynamics with fermions $f = e,\mu\ldots$~,
knowledge of the vacuum polarization spectral
function determined from the tree level $e^{+}e^{-}\ra f^{+}f^{-}$
cross sections, together with a single low energy measurement of
the fine structure constant $\alpha$, enables the construction of the 
one--loop effective charge $\alpha_{\rm eff}(q^2)$ for all $q^2$. 
It is shown how an identical
procedure can be followed in the electroweak sector of the Standard Model to
construct three gauge--, scale-- and scheme--independent 
one--loop electroweak effective charges and an effective weak mixing angle 
from the tree level  
$e^{+}e^{-}\ra W^{+}W^{-},\, ZH$ and 
$e^{+} \nu_{e}\ra W^{+}Z, \,W^{+}\gamma,\,W^{+}H$
differential cross sections, together with three low energy measurements,
which may be chosen to be $\alpha$ and the masses of the $W$ and $Z$ bosons.
It is found that the corresponding proper self--energy--like functions
thus constructed are 
{\em identical} to those obtained in the pinch technique framework. 
In this way, it is shown how the concept of effective charges in the 
electroweak Standard Model is as well--defined 
and unique as in quantum electrodynamics.
 
\end{abstract}
 
\vfill
 
%\begin{flushleft} hep-ph/96...\\
%\end{flushleft}
 
\end{titlepage}
 
%%%%%%%%%%%%%%%%%%%%%%%%%%%%%%%%%%%%%%%%%%%%%%%%
%%%%%%%%%%%%%%%%%%%%%%%%%%%%%%%%%%%%%%%%%%%%%%%%

\section {Introduction}
 
The possibility of extending the concept of
an effective charge \cite{gellmannlow} from quantum electrodynamics
to non--abelian gauge theories
is of fundamental interest for at least three reasons.
First, in quantum chromodynamics, the existence of
an effective charge analogous
to that of QED is explicitly assumed in renormalon analyses 
of the behaviour of pertubation series at high orders \cite{renormalon}.
The ability to identify directly and unambiguously the
infinite subset of gluon self--energy--like radiative corrections that
one is summing in such analyses
is important in order to provide a well--defined
basis for renormalon calculus.
Second, in theories involving unstable particles,
e.g.\ the Standard Model, the presence of such particles
necessarily requires the Dyson summation of infinite subsets of
radiative corrections in order to regulate the
singularities which otherwise occur in the corresponding
tree level propagators \cite{unstable}. Such summations,
directly linked to the concept of an effective charge,
are essential for the evaluation
of physical amplitudes at arbitrary values of the kinematic parameters.
Third, in theories involving disparate energy scales,
e.g.\ grand unified theories,
the extraction of accurate low--energy predictions
requires an exact treatment of threshold effects due to heavy particles
\cite{threshold}. The ability to construct a set of effective charges,
valid for all momenta $q^{2}$ and
not just the asymptotic regime governed by the
renormalization group $\beta$--functions, would automatically
provide the natural way to account for such threshold effects.
In all cases, the fundamental problem is the gauge dependence,
and hence ambiguity,
of the gauge boson self--energies in a non--abelian gauge theory.
This gauge dependence necessarily
means that the gauge boson self--energies in
such theories are not directly related to measurable quantities.
 
Recently however, there has been substantial progress in understanding
how the QED concept of an effective charge can be extended
to non--abelian gauge theories \cite{njw1,papapil1,papapil2,njw2}.
The theoretical framework which has enabled this progress
is the pinch technique (PT) 
\cite{cornwall1,cornwall2,PTfirst,PTmassive,degsir,njw3,PTrest}.
The PT is a well--defined algorithm for the rearrangement of conventional
gauge--dependent one--loop $n$--point functions
to construct individually gauge--independent one--loop
``effective'' $n$--point functions. This rearrangement of perturbation
theory is based on the systematic use of the tree level Ward identities
of the theory to cancel in Feynman amplitudes all factors of longitudinal
four--momentum associated with gauge fields propagating in loops.
In addition to being gauge--independent, the PT ``effective'' $n$--point
functions display many desirable theoretical properties. In particular,
they obey the same Ward identities as the corresponding tree level
functions.
 
The purpose of this paper is to discuss the relationship
between effective charges and the cross sections for
certain physical processes in the electroweak sector of the Standard Model.
It is well known that in QED with fermions $f = e,\mu\ldots$~,
the imaginary part of, e.g., the muon contribution to the one--loop
vacuum polarization is directly related to the tree level
cross section for the process $e^{+}e^{-} \ra \mu^{+}\mu^{-}$.
This relation is a result purely of the unitarity of the S--matrix
\hfill $S = I + iT$, \hfill expressed \hfill in \hfill the \hfill optical 
\hfill theorem \hfill for \hfill the \hfill particular \hfill case 
\hfill of 
 
\pagebreak
 
\begin{center}
\begin{picture}(400,130)(0,370)
 
\put(-10,450){\makebox(0,0)[c]{\Large Im}}
\put( 12,414){\makebox(0,0)[c]{\small $e^{+}$}}
\put( 12,493){\makebox(0,0)[c]{\small $e^{-}$}}
\ArrowLine( 20,420)(35,450)
\ArrowLine( 20,480)(35,450)
\Photon( 35,450)( 62.5,450){4}{2.5}
\ArrowArc( 80,450)(17.5,45,405)
\put( 60,430){\makebox(0,0)[c]{\small $\mu$}}
\DashLine( 80,410)( 80,490){5}
\Photon( 97.5,450)(125,450){-4}{2.5}
\ArrowLine(125,450)(140,420)
\ArrowLine(125,450)(140,480)
\put(150,414){\makebox(0,0)[c]{\small $e^{+}$}}
\put(150,493){\makebox(0,0)[c]{\small $e^{-}$}}
 
\put(200,450){\makebox(0,0)[c]{\LARGE $=$}}
 
\Line(265,410)(265,490)
\Line(375,410)(375,490)
\put(390,495){\makebox(0,0)[c]{\Large 2}}
\put(287,414){\makebox(0,0)[c]{\small $e^{+}$}}
\put(287,493){\makebox(0,0)[c]{\small $e^{-}$}}
\ArrowLine(290,420)(305,450)
\ArrowLine(290,480)(305,450)
\Photon(305,450)(335,450){4}{3}
\ArrowLine(335,450)(350,420)
\ArrowLine(335,450)(350,480)
\put(360,414){\makebox(0,0)[c]{\small $\mu^{+}$}}
\put(360,493){\makebox(0,0)[c]{\small $\mu^{-}$}}
 
%\put(200,360){\makebox(0,0)[c]{\Large (a)}}
 
\end{picture}
\end{center}
 
\noindent
{\footnotesize Fig.~1.
The relation between the imaginary part of the
muon contribution to the one-loop vacuum polarization and the
tree level cross section
$\sigma(e^{+}e^{-}\rightarrow \mu^{+}\mu^{-})$ in QED.}

\vspace{20pt}
 
\noi
forward scattering in the process $e^{+}e^{-} \ra e^{+}e^{-}$ :
\be\label{opth}
\im \langle e^{+}e^{-}|T|e^{+}e^{-} \rangle =
\frac{1}{2}\sum_{i}\int d\Gamma_{i}\,|\langle e^{+}e^{-}|T|i\rangle |^{2}\,.
\ee
In \eq{opth}, the sum on the right hand side is over all
on--shell physical states
$|i\rangle$ compatible with the quantum numbers of $|e^{+}e^{-}\rangle$;
in each case the integral is over the available phase space $\Gamma_{i}$.
The tree level contribution of the muon pair $|\mu^{+}\mu^{-}\rangle$
to the r.h.s.\ of \eq{opth} and the corresponding imaginary part of
the muon contribution to the one--loop vacuum polarization on the
l.h.s.\ of \eq{opth} are illustrated schematically in Fig.~1.
The muon contribution to the renormalized one--loop vacuum polarization
may thus be reconstructed directly from the tree level cross section
$\sigma(e^{+}e^{-} \ra \mu^{+}\mu^{-})$
via a once--subtracted dispersion relation. Similarly, the
tree level contributions of the electron--positron pair
$|e^{+}e^{-}\rangle$
to the r.h.s.\ of \eq{opth} and the corresponding imaginary parts of
the electron contribution to the one--loop diagrams on the
l.h.s.\ of \eq{opth} are illustrated schematically in Fig.~2.
As indicated explicitly in Fig.~2,
the imaginary part of the one--loop self--energy [box]
diagram on the l.h.s.\ corresponds
to the direct contribution of the $s$--channel [$t$--channel]
tree level photon exchange diagram on the r.h.s.,
while the imaginary part of the one--loop vertex diagram on the
l.h.s.\ corresponds to the interference contribution of the
tree level diagrams on the r.h.s.
The electron contribution to the vacuum polarization may thus
be reconstructed directly from the self--energy--like component\footnote{
The self--energy--like component of the experimental
Bhabha cross section may be projected out from the
full cross section using a procedure analogous to that to be described
here in Sec.~8.}
of the tree level Bhahba scattering cross section
$\sigma(e^{+}e^{-} \ra e^{+}e^{-})$ again
via a once--subtracted dispersion relation.
Knowledge of each such contribution to the vacuum polarization,
together with a single low energy measurement of the fine structure
constant $\alpha = 1/137.036\ldots$~, then enables the QED
one--loop effective charge $\alpha_{\rm eff}(q^{2})$ to be uniquely
constructed for all values of its argument.

\begin{center}
\begin{picture}(400,330)(0,-30)

%%%%%%%%%%%%%%%%%%%%%%%%%%%%%%%%%%%%%%%%%%%%%
 
\put(-10,250){\makebox(0,0)[c]{\Large Im}}
\put( 12,214){\makebox(0,0)[c]{\small $e^{+}$}}
\put( 12,293){\makebox(0,0)[c]{\small $e^{-}$}}
\ArrowLine( 20,220)(35,250)
\ArrowLine( 20,280)(35,250)
\Photon( 35,250)( 62,250){4}{2.5}
\ArrowArc( 80,250)(17.5,45,405)
\put( 60,230){\makebox(0,0)[c]{\small $e$}}
\DashLine( 80,210)( 80,290){5}
\Photon( 98,250)(125,250){-4}{2.5}
\ArrowLine(125,250)(140,220)
\ArrowLine(125,250)(140,280)
\put(150,214){\makebox(0,0)[c]{\small $e^{+}$}}
\put(150,293){\makebox(0,0)[c]{\small $e^{-}$}}
 
\put(200,250){\makebox(0,0)[c]{\LARGE $=$}}
 
\Line(265,210)(265,290)
\Line(375,210)(375,290)
\put(390,295){\makebox(0,0)[c]{\Large 2}}
\put(287,214){\makebox(0,0)[c]{\small $e^{+}$}}
\put(287,293){\makebox(0,0)[c]{\small $e^{-}$}}
\ArrowLine(290,220)(305,250)
\ArrowLine(290,280)(305,250)
\Photon(305,250)(335,250){4}{3}
\ArrowLine(335,250)(350,220)
\ArrowLine(335,250)(350,280)
\put(360,214){\makebox(0,0)[c]{\small $e^{+}$}}
\put(360,293){\makebox(0,0)[c]{\small $e^{-}$}}
 
%%%%%%%%%%
 
\put(-10,150){\makebox(0,0)[c]{\Large Im}}
\ArrowLine( 20,120)(35,150)
\ArrowLine( 20,180)(35,150)
\Photon( 35,150)( 65,150){4}{3}
\DashLine( 95,110)( 95,190){5}
\Line( 65,150)(110,132)
\Line( 65,150)(110,168)
\Photon(110,132)(110,168){-4}{3.5}
\ArrowLine(110,132)(140,120)
\ArrowLine(110,168)(140,180)
 
\put(200,150){\makebox(0,0)[c]{\LARGE $=$}}
 
\Line(223,110)(223,190)
\Line(317,110)(317,190)
\Line(323,110)(323,190)
\Line(417,110)(417,190)
\ArrowLine(240,120)(255,150)
\ArrowLine(240,180)(255,150)
\Photon(255,150)(285,150){4}{3}
\ArrowLine(285,150)(300,120)
\ArrowLine(285,150)(300,180)
\ArrowLine(340,120)(370,135)
\ArrowLine(340,180)(370,165)
\Photon(370,135)(370,165){4}{3}
\ArrowLine(370,135)(400,120)
\ArrowLine(370,165)(400,180)
 
%%%%%%%%%%
 
\put(-10, 50){\makebox(0,0)[c]{\Large Im}}
\ArrowLine( 20, 20)(50,32)
\ArrowLine( 20, 80)(50,68)
\Photon( 50, 32)( 50, 68){4}{3.5}
\Line( 50, 32)(110,32)
\Line( 50, 68)(110,68)
\DashLine( 80,10)( 80, 90){5}
\Photon(110, 32)(110, 68){-4}{3.5}
\ArrowLine(110, 32)(140, 20)
\ArrowLine(110, 68)(140, 80)
 
\put(200, 50){\makebox(0,0)[c]{\LARGE $=$}}
 
\Line(265, 10)(265, 90)
\Line(375, 10)(375, 90)
\put(390, 95){\makebox(0,0)[c]{\Large 2}}
\ArrowLine(290, 20)(320, 35)
\ArrowLine(290, 80)(320, 65)
\Photon(320, 35)(320, 65){4}{3}
\ArrowLine(320, 35)(350, 20)
\ArrowLine(320, 65)(350, 80)
 
%\put(200,-40){\makebox(0,0)[c]{\Large (b)}}
 
\end{picture}
\end{center}
 
\noindent
{\footnotesize Fig.~2.
The relation between the imaginary parts of the
electron contribution to the one--loop vacuum polarization,
vertex and box diagrams and the components of the
tree level cross section
$\sigma(e^{+}e^{-}\rightarrow e^{+}e^{-})$ in QED.}

\vspace{20pt}
 
When QED is embedded in the electroweak sector of the Standard Model,
there occur further lowest order contributions to the relation \eq{opth}.
The tree level contributions of the gauge boson pair $|W^{+}W^{-}\rangle$
to the r.h.s.\ of \eq{opth} and the corresponding imaginary parts
of the contributions of
the $W$ and its associated would--be Goldstone boson
and ghost to the one--loop diagrams
for $e^{+}e^{-}\ra e^{+}e^{-}$ on the l.h.s.\ of \eq{opth}
are illustrated schematically in Fig.~3 (for simplicity,
the coupling of the Higgs to $e^{+}e^{-}$ has been neglected).
There are two basic and well--known observations to make
regarding the equation illustrated in Fig.~3.
First, on the r.h.s.\ of Fig.~3,
each of the contributions corresponding to the individual
tree level diagrams violates unitarity \cite{allbowbur}.
It is only when the contributions, direct plus interference, from
all three diagrams are combined that,
as a result of extensive cancellations, the overall contribution on the
r.h.s.\ of Fig.~3 is well--behaved at high energies.
Second, on the l.h.s.\ of Fig.~3,
each of the contributions due to the
imaginary parts of the one--loop self--energy, vertex and box diagrams
is individually gauge--dependent \cite{fujleesan}.
It is only when the contributions are
combined that the gauge dependencies, in particular
the unphysical thresholds that in general
 
\pagebreak
 
\begin{center}
\begin{picture}(400,340)(0,-40)
 
%\put(0,-20){\framebox(400,320){}}
 
\put( 44,212){\makebox(0,0)[c]{\small $e^{+}$}}
\put( 44,293){\makebox(0,0)[c]{\small $e^{-}$}}
\ArrowLine( 50,220)(65,250)
\ArrowLine( 50,280)(65,250)
\put( 80,265){\makebox(0,0)[c]{\small $\gamma,Z$}}
\Photon( 65,250)( 92,250){4}{2.5}
\CArc(110,250)(17.5,0,360)
\CArc(110,250)(17  ,0,360)
\CArc(110,250)(16.5,0,360)
\CArc(110,250)(16  ,0,360)
\CArc(110,250)(15.5,0,360)
\put(111.5,251){\makebox(0,0)[c]{\Large $W$}}
\DashLine(110,210)(110,242){5}
\DashLine(110,258)(110,290){5}
\put(140,265){\makebox(0,0)[c]{\small $\gamma,Z$}}
\Photon(128,250)(155,250){-4}{2.5}
\ArrowLine(155,250)(170,220)
\ArrowLine(155,250)(170,280)
\put(177,212){\makebox(0,0)[c]{\small $e^{+}$}}
\put(177,293){\makebox(0,0)[c]{\small $e^{-}$}}

\put( 44,112){\makebox(0,0)[c]{\small $e^{+}$}}
\put( 44,193){\makebox(0,0)[c]{\small $e^{-}$}}
\ArrowLine( 50,120)(65,150)
\ArrowLine( 50,180)(65,150)
\put( 80,165){\makebox(0,0)[c]{\small $\gamma,Z$}}
\Photon( 65,150)( 95,150){4}{3}
\put(126.5,150){\makebox(0,0)[c]{\Large $W$}}
\DashLine(125,110)(125,142){5}
\DashLine(125,158)(125,190){5}
\Line( 95.16,150)(140,132)
\Line( 96.32,150)(140,132.33)
\Line( 97.48,150)(140,132.66)
\Line( 98.64,150)(140,133)
\Line( 99.80,150)(140,133.33)
\Line( 95.16,150)(140,168)
\Line( 96.32,150)(140,167.66)
\Line( 97.48,150)(140,167.33)
\Line( 98.64,150)(140,167)
\Line( 99.80,150)(140,166.66)
\Line(140,132)(140,168)
\Line(139.5,132.2)(139.5,167.8)
\Line(139,132.4)(139,167.6)
\Line(138.5,133.6)(138.5,166.4)
\Line(138,133.8)(138,166.2)
\ArrowLine(140,132)(170,120)
\ArrowLine(140,168)(170,180)
\put(177,112){\makebox(0,0)[c]{\small $e^{+}$}}
\put(177,193){\makebox(0,0)[c]{\small $e^{-}$}}
\put( 90,115){\makebox(0,0)[c]{\small +\,\,reversed}}

\put( 44, 12){\makebox(0,0)[c]{\small $e^{+}$}}
\put( 44, 93){\makebox(0,0)[c]{\small $e^{-}$}}
\ArrowLine( 50, 20)(80,32)
\ArrowLine( 50, 80)(80,68)
\Boxc(110,50)(60,36)
\Boxc(110,50)(59,35)
\Boxc(110,50)(58,34)
\Boxc(110,50)(57,33)
\Boxc(110,50)(56,33)
\put(111.5, 50){\makebox(0,0)[c]{\Large $W$}}
\DashLine(110, 10)(110, 42){5}
\DashLine(110, 58)(110, 90){5}
\ArrowLine(140, 32)(170, 20)
\ArrowLine(140, 68)(170, 80)
\put(177, 12){\makebox(0,0)[c]{\small $e^{+}$}}
\put(177, 93){\makebox(0,0)[c]{\small $e^{-}$}}

\put(-10,150){\makebox(0,0)[c]{\Large Im}}
\put( 20,150){\makebox(0,0)[c]
{$ \left\{ \begin{array}{c}
\\ \\ \\ \\ \\ \\ \\ \\ \\ \\ \\ \\ \\ \\ \\ \\ \\ \\ \\
\end{array} \right.$}}
\put(200,150){\makebox(0,0)[c]
{$ \left. \begin{array}{c}
\\ \\ \\ \\ \\ \\ \\ \\ \\ \\ \\ \\ \\ \\ \\ \\ \\ \\ \\
\end{array} \right\}$}}
\put(235,150){\makebox(0,0)[c]{\LARGE $=$}}

\Line(265, 50)(265,250)
\Line(395, 50)(395,250)
\put(410,255){\makebox(0,0)[c]{\Large 2}}
 
\put(294,162){\makebox(0,0)[c]{\small $e^{+}$}}
\put(294,243){\makebox(0,0)[c]{\small $e^{-}$}}
\ArrowLine(300,170)(315,200)
\ArrowLine(300,230)(315,200)
\put(330,215){\makebox(0,0)[c]{\small $\gamma,Z$}}
\Photon(315,200)(345,200){4}{3}
\Photon(345,200)(360,170){4}{3}
\ArrowLine(353,185)(350.5,182)
\Photon(345,200)(360,230){4}{3}
\ArrowLine(351,215)(356,215)
\put(371,162){\makebox(0,0)[c]{\small $W^{+}$}}
\put(371,242){\makebox(0,0)[c]{\small $W^{-}$}}
 
\put(294, 62){\makebox(0,0)[c]{\small $e^{+}$}}
\put(294,143){\makebox(0,0)[c]{\small $e^{-}$}}
\ArrowLine(300, 70)(330, 85)
\ArrowLine(300,130)(330,115)
\put(322,100){\makebox(0,0)[c]{\small $\nu_{e}$}}
\Line(330, 85)(330,115)
\Photon(330, 85)(360, 70){-4}{3}
\ArrowLine(345, 77.5)(347, 79)
\Photon(330,115)(360,130){4}{3}
\ArrowLine(345,122.5)(347,121)
\put(371, 62){\makebox(0,0)[c]{\small $W^{+}$}}
\put(371,142){\makebox(0,0)[c]{\small $W^{-}$}}
 
\end{picture}
\end{center}
 
\noindent
{\footnotesize Fig.~3.
The schematic representation of the tree level contribution of the
gauge boson pair $|W^{+}W^{-}\rangle$ to the r.h.s.\ of
\eq{opth} and the corresponding imaginary parts of the
contributions of the $W$ and its associated would--be Goldstone boson
and ghost, collectively denoted by ``$W$'',
to the one--loop self-energy, vertex and box diagrams for the process
$e^{+}e^{-}\rightarrow e^{+}e^{-}$ on the l.h.s.\ of \eq{opth}
in the electroweak Standard Model.}
 
\vspace{20pt}
 
\noi
occur, cancel and the
overall contribution on the l.h.s.\ of Fig.~3 is gauge--independent.
Either of these facts alone is sufficient to prevent
the unitarity relation illustrated in Fig.~3 from
holding for the {\em individual} diagrammatic contributions
which occur on each side. Thus, the simple QED relation between
the components of the tree level cross sections for
the interaction of on--shell particles and the imaginary parts of
the corresponding one--loop self--energy, vertex and box
diagrams is apparently lost.
 
In this paper, it is shown how the QED--like correspondence between individual
contributions on each side of \eq{opth} may be maintained in the
electroweak Standard Model. The starting point is a simple
re--analysis of the calculation of the tree level cross section for
the process $e^{+}e^{-} \ra W^{+}W^{-}$.
The crucial observation is that
the cancellation mechanism responsible for the good
high energy behaviour of the overall cross section
occurs directly at the level of the tree level Feynman diagrams.
Implementing this cancellation at the very first step in the calculation,
rather than the very last,
the cross section is shown to decompose naturally into
components which are individually well--behaved at high energy.
Then, by direct analogy with QED,
the resulting self--energy--like cross section components
may be used to {\em define} manifestly gauge--independent
$W^{+}W^{-}$ contributions to photon and $Z$ one--loop
renormalized self--energies via dispersion relations.
These self--energy contributions
are found to be identical to those obtained in the pinch technique.
It is then shown how these self--energies,
together with that for the $W$ gauge boson,
define three gauge--independent
electroweak effective charges and an effective weak mixing angle.
These four effective quantities are shown to be renormalization
scale-- and scheme--independent, and at high energies to match on to the
corresponding running quantities defined from the two electroweak
$\beta$--functions.
Finally, it is described how the self--energy--like components
of the $e^{+}e^{-} \ra W^{+}W^{-}$ cross section
used to construct the $W^{+}W^{-}$ contributions to the
photon and $Z$ self--energies
may be extracted directly from experiment. Explicit expressions
are given for the required projection functions in the simplified
case in which the weak mixing angle is set to zero.
In this way, it is shown how the concept of an effective
charge in the electroweak Standard Model is as 
well--defined and unique as in QED.

The paper is organized as follows.
In Sec.~2, a brief review is given of properties of
the vacuum polarization and effective charge in QED.
In Sec.~3, the re--analysis of the calculation of the tree level
cross section for $e^{+}e^{-}\ra W^{+}W^{-}$ is presented,
together with that for $e^{+}e^{-}\ra ZH$,
where $H$ is the Higgs boson.
In Sec.~4, the definitions are given of the
$W$ and $ZH$ contributions to photon and $Z$
one--loop self--energies in terms of the cross section components.
In Sec.~5, the corresponding set of effective charges and the effective
weak mixing angle are constructed.
In Sec.~6, it is shown how these four effective quantities are related
to the two running couplings defined from the renormalization group.
In Sec.~7, a comparison is made between the effective charges
constructed here and those obtained in the background field method.
In Sec.~8, the extraction of the required
cross section components from experiment is described.
The paper finishes with our conclusions in Sec.~9.
Some technical details are relegated to three Appendices.
Throughout, we consider only gauge field contributions to one--loop
self--energies, since fermion and scalar contributions are
standard.\footnote{For a recent extensive discussion of fermion loop
effects in $e^{+}e^{-}$ annihilation, see Ref.~\cite{beenaker}.}

%%%%%%%%%%%%%%%%%%%%%%%%%%%%%%%%%%%%%%%%%%%%%%%%

\setcounter{equation}{0}
 
\section{The Effective Charge in QED}
 
We first review some basic properties of the vacuum polarization
and effective charge in QED with fermions $f = e,\mu\ldots$~.
For more details see, e.g., Refs.~\cite{books}.
 
i) The vacuum polarization function
$\Pi(q^{2})$ is gauge--independent at all $q^{2}$ and to all orders
in perturbation theory.
 
ii) At the one--loop level, the imaginary part of $\Pi(q^{2})$
is directly related, via the optical theorem, to the tree level
cross sections for the physical processes
$e^{+}e^{-}\rightarrow f^{+}f^{-}$. 
The differential cross section for these processes is given by
\bea
\lefteqn{
\frac{d\sigma(e^{+}e^{-}\rightarrow f^{+}f^{-})}{d\Omega}
\,\,\,=\,\,\,
\frac{\alpha^{2}}{4s}\Biggl\{ 
\frac{\beta_{f}}{\beta_{e}}\biggl[
\,3-\beta_{e}^{2}-\beta_{f}^{2}+\beta_{e}^{2}\beta_{f}^{2}\cos^{2}\theta\, 
\biggr] }\nn \\
& &
\mbox{}+\delta_{ef}\frac{1}{(1-\cos\theta)}\,\frac{1}{\beta_{e}^{2}}\biggl[
-6 + 4\beta_{e}^{2} - 4\beta_{e}^{2}(2-\beta_{e}^{2})\cos\theta 
- 2\beta_{e}^{4}\cos^{2}\theta\, \biggr] \nn \\
& &
\mbox{}+\delta_{ef}\frac{1}{(1-\cos\theta)^{2}}\,\frac{1}{\beta_{e}^{4}}\biggl[
\,\,4 + 6\beta_{e}^{2} + 4\beta_{e}^{2}(2-\beta_{e}^{2})\cos\theta 
+ 2\beta_{e}^{4}\cos^{2}\theta\, \biggr]
\Biggr\}\vartheta(s - 4m_{f}^{2})\,,
\label{dsigmaff}
\eea
where $\alpha = e^{2}/4\pi$,
$s$ is the square of the total centre of mass energy,
$\theta$ is the centre of mass scattering angle and
$\beta_{f} = \sqrt{1 - 4m_{f}^{2}/s}$~,
where $m_{f}$ is the mass of the fermion $f$.
In \eq{dsigmaff}, the terms grouped in the first, second and third sets of
square parentheses correspond to the self--energy--,
vertex-- and box--like contributions, respectively, shown in Fig.~2.
The fact that the vertex-- and box--like contributions only occur
in the case $f = e$, i.e.\ Bhabha scattering, is indicated by the
Kronecker deltas $\delta_{ef}$. From the relation \eq{opth},
the imaginary part of the one--loop contribution of the virtual fermion $f$
to $\Pi(q^{2})$ is then given directly by the self--energy--like component
of the tree level cross section $\sigma(e^{+}e^{-}\ra f^{+}f^{-})$:
\bea\label{sigmaff1}
e^{2}\frac{1}{s}\,\im \Pi^{(ff)}(s) &=& \frac{2\beta_{e}}{(3-\beta_{e}^{2})}
\,\sigma_{\rm s.e.l.}(e^{+}e^{-}\ra f^{+}f^{-}) \\
\label{sigmaff2}
&=& 2\pi\alpha^{2}\frac{1}{s}\beta_{f}
\Biggl\{\frac{2}{3} + \frac{4m_{f}^{2}}{3s}\Biggr\}
\vartheta(s - 4m_{f}^{2})\,,
\eea
where the subscript ``s.e.l.'' denotes ``self--energy--like''. 

iii) Given a particular contribution to the spectral function $\im \Pi(s)$,
the corresponding contribution to the
renormalized vacuum polarization function
$\Pi_{R}(q^2)$ can be reconstructed via a once--subtracted
dispersion relation. For example, for the one--loop contribution
of the fermion $f$, choosing the on--shell renormalization scheme,
\bea\label{qeddisprel}
\Pi_{R}^{(ff)}(q^2) &=&
q^{2}\,\int_{4m_{f}^{2}}^{\infty}ds\frac{1}{s(s-q^2)}
\frac{1}{\pi}\,\im \Pi^{(ff)}(s) \\ &=&
-\frac{\alpha}{\pi}\times
\left\{\begin{array}{ll} {\displaystyle
-\frac{1}{15}\frac{q^2}{m_{f}^{2}}
+\cO\biggl(\frac{q^{4}}{m_{f}^{4}}\biggr) }
& \,\,\,\, \phantom{-}q^{2}/m_{f}^{2} \rightarrow 0 \\
{\displaystyle
+\frac{2}{3}\,\frac{1}{2}\ln\biggl(-\frac{q^2}{m_{f}^{2}}\biggr)
-\frac{5}{9} +\cO\biggl(\frac{m_{f}^{2}}{q^2}\biggr) } &
\,\,\,\,-q^{2}/m_{f}^{2} \rightarrow \infty\,,
\end{array}
\right.
\label{PiRffQED}
\eea
The coefficient +2/3 of the logarithmic term in \eq{PiRffQED},
originating from the +2/3 in the parentheses in \eq{sigmaff2},
is precisely the contribution of the fermion
to the first coefficient $\beta_{1}$ of the QED $\beta$--function.
 
iv) The infinite subset of radiative corrections summed in
the Dyson series generated by the
one--particle--irreducible vacuum polarization $\Pi_{R}(q^{2})$
defines an {\it effective charge}
which is gauge--, scale--, and scheme--independent
to all orders in perturbation theory:
\be\label{QEDalphaeff}
\alpha_{\rm eff}(q^{2}) =
\frac{e_{R}^{2}}{4\pi}\,\frac{1}{1 + \Pi_{R}(q^{2})} =
\frac{e^{2}}{4\pi}\,\frac{1}{1 + \Pi(q^{2})}\,,
\ee
where we have used $e_{R}^{2} = (Z_{2}^{2}Z_{3}/Z_{1}^{2})e^{2}$ and $1 +
\Pi_{R} = Z_{3}(1 + \Pi)$
together with the QED Ward identity $Z_{1} = Z_{2}$ to write
$\alpha_{\rm eff}(q^{2})$ purely in terms of bare quantities.

v) At $-q^{2}/m_{f}^{2} \ra\infty$, the effective charge
$\alpha_{\rm eff}(q^{2})$ matches on to the {\em running coupling}
$\alphab(q^{2})$ defined from the renormalization group:
at the one--loop level,
\be
\lim_{-q^{2}/m_{f}^{2}\ra\infty} \alpha_{\rm eff}(q^2) = \alphab(q^{2}) = 
\frac{\alpha_{R}}
{1 - \frac{\alpha_{R}}{\pi}\beta_{1}\frac{1}{2}\log(-q^{2}/m_{f}^{2})}\,,
\ee
where $\beta_{1} = +\frac{2}{3}n_{f}$ for $n_{f}$ species of fermion.

vi) At $q^{2} = 0$, $\Pi(0)$ specifies fully the one--loop radiative
corrections to the tree level Compton scattering process
$\gamma e^{-} \ra \gamma e^{-}$ in the limit of vanishing photon energy.
The fine structure constant $\alpha = 1/137.036\ldots $
which governs Compton scattering in the classical Thomson limit of
vanishing photon energy then provides the low--energy definition
of the coupling constant for the quantum theory: at the one--loop level,
\be
\alpha =
\frac{e^{2}}{4\pi}\biggl( 1 - \Pi(0) \biggr) =
\frac{e_{R}^{2}}{4\pi}\biggl( 1 - \Pi_{R}(0) + {\cal O}(e^{4})\biggr)\,.
\ee
In the on--shell scheme, $\Pi_{R}(0) = 0$ and
$e_{R}$ is identified with the physical electron charge.
At $q^{2} = 0$, the effective charge therefore matches on to the
fine structure constant: $\alpha_{\rm eff}(0) = \alpha$.

vii) For $f\neq e$, the imaginary part of the one--loop
contribution of the fermion $f$ to the spectral function $\im \Pi(s)$
is measured directly in the the tree level cross section for
$e^{+}e^{-} \ra f^{+}f^{-}$.
For $f = e$, it is necessary to isolate the self--energy--like
component of the tree level Bhabha cross section. As a result of the
form of the numerators in the second and third square parentheses
in \eq{dsigmaff}, the decomposition 
of the Bhabha differential cross section 
into self--energy--, vertex-- and box--like components
does {\em not} correspond to some set of simple kinematic criteria.
The three components of the differential cross section 
\eq{dsigmaff} are however
linearly independent functions of $\cos\theta$.
They may therefore each be projected out from the measured cross section
by convolution with
appropriately chosen polynomials in $\cos\theta$.

Thus, in QED, knowledge of the spectral function $\im \Pi(s)$,
determined from the tree level $e^{+}e^{-} \ra f^{+}f^{-}$ cross sections,
together with a single low energy measurement of the fine structure constant
$\alpha$
(obtained e.g.\ from the ac Josephson effect and the quantized Hall effect, 
or from the anomalous magnetic moment of the electron~\cite{alphaval}), 
enables the construction of the one--loop effective charge
$\alpha_{\rm eff}(q^2)$ for all $q^2$. 
In the remainder of this paper, it will be shown how an identical
procedure can be followed in the electroweak sector of the Standard Model
to construct a set of gauge--, scale-- and scheme--independent
electroweak effective charges.
 
%%%%%%%%%%%%%%%%%%%%%%%%%%%%%%%%%%%%%%%%%%%%%%%%%
 
\setcounter{equation}{0}
 
\section{The Tree Level Processes $e^{+}e^{-} \rightarrow
W^{+}W^{-}$ and $e^{+}e^{-} \rightarrow ZH$}
 
In the electroweak sector of the Standard Model,
the on--shell states involving the electroweak gauge bosons
$W$ and $Z$ which contribute to the r.h.s.\ of \eq{opth}
at lowest order are $|W^{+}W^{-}\rangle$ and $|ZH\rangle$,
corresponding to the tree level physical processes
$e^{+}e^{-} \rightarrow W^{+}W^{-}$ and
$e^{+}e^{-} \rightarrow ZH$, where $H$ is the Higgs boson. We shall discuss
these processes separately.
 
%%%%%%%%%%%%%%%%%%%%%%%
 
\begin{center}
\begin{picture}(400,130)(0,20)
 
%\put(0, 20){\framebox(400,180){}}

\put( 34, 58){\makebox(0,0)[c]{\small $e^{+}(k_{2},s_{2})$}}
\put( 34,144){\makebox(0,0)[c]{\small $e^{-}(k_{1},s_{1})$}}
\ArrowLine( 40, 70)( 55,100)
\ArrowLine( 40,130)( 55,100)
\put( 73,115){\makebox(0,0)[c]{\small $\gamma$}}
\Photon( 55,100)( 85,100){4}{3}
\Photon( 85,100)(100, 70){4}{3}
\ArrowLine( 93, 85)( 90.5, 82)
\Photon( 85,100)(100,130){4}{3}
\ArrowLine( 91,115)( 96,115)
\put(111, 58){\makebox(0,0)[c]{\small $W^{+}(p_{2},\lambda_{2})$}}
\put(111,144){\makebox(0,0)[c]{\small $W^{-}(p_{1},\lambda_{1})$}}
\put( 70,40){\makebox(0,0)[c]{(a)}}

\ArrowLine(170, 70)(185,100)
\ArrowLine(170,130)(185,100)
\put(203,115){\makebox(0,0)[c]{\small $Z$}}
\Photon(185,100)(215,100){4}{3}
\Photon(215,100)(230, 70){4}{3}
\ArrowLine(223, 85)(220.5, 82)
\Photon(215,100)(230,130){4}{3}
\ArrowLine(221,115)(226,115)
\put(200,40){\makebox(0,0)[c]{(b)}}
 
\ArrowLine(300, 70)(330, 85)
\ArrowLine(300,130)(330,115)
\put(322,100){\makebox(0,0)[c]{\small $\nu_{e}$}}
\Line(330, 85)(330,115)
\Photon(330, 85)(360, 70){-4}{3}
\ArrowLine(345, 77.5)(347, 79)
\Photon(330,115)(360,130){4}{3}
\ArrowLine(345,122.5)(347,121)
\put(330,40){\makebox(0,0)[c]{(c)}}
 
\end{picture}
\end{center}
 
\noindent
{\small Fig.~4.
The three diagrams which contribute to the process
$e^{+}e^{-} \rightarrow W^{+}W^{-}$ at tree level for the case
of massless fermions.}
 
%\vspace{10pt}
\vspace{5pt}

\subsection{$e^{+}e^{-} \rightarrow W^{+}W^{-}$}
 
We first consider the process
$e^{-}(k_{1},s_{1})e^{+}(k_{2},s_{2})
\rightarrow W^{-}(p_{1},\lambda_{1})W^{+}(p_{2},\lambda_{2})$,
where $s_{1},s_{2}$ and $\lambda_{1},\lambda_{2}$ 
label the polarizations of the on--shell initial and final states.
For simplicity, we neglect the fermion
mass ($m_{e} = 0$), and hence the contribution to this process due
to Higgs exchange. The three diagrams which contribute at tree level are then
as shown in Fig.~4. The relevant kinematic variables are
\bea
s &=& (k_{1}+k_{2})^{2} \,\,\,=\,\,\, (p_{1}+p_{2})^{2}\,, \label{s}\\
t &=& (k_{1}-p_{1})^{2} \,\,\,=\,\,\, (p_{2}-k_{2})^{2}
\,\,\,=\,\,\,-\frac{1}{4}s\,(1 + \beta^{2} - 2\beta\cos\theta)\,, \label{t}
\eea
where
\be
\beta = \sqrt{1 - \frac{4M_{W}^{2}}{s}}
\ee
and $\theta$ is the centre of mass scattering angle.
 
The S--matrix element for this process is given by
\be\label{T}
i\langle W^{+}W^{-}|T|e^{+}e^{-}\rangle 
=
i\epsilon_{\mu}^{*}(p_{1},\lambda_{1})\epsilon_{\nu}^{*}(p_{2},\lambda_{2})
\,\overline{v}(k_{2},s_{2})T^{\mu\nu}u(k_{1},s_{1})\,,
\ee
where $\epsilon_{\mu}^{*}$, $\epsilon_{\nu}^{*}$ are the
$W^{-}$, $W^{+}$ gauge boson polarization vectors, respectively,
$u$ and $\overline{v}$ are the fermion spinors and the amputated
Green's function $T^{\mu\nu}$ is given by the sum of the
diagrams in Fig.~4.
The square of the modulus of $\langle W^{+}W^{-}|T|e^{+}e^{-}\rangle$,
averaged over the initial state polarizations
and summed over the final state polarizations, may be written
\bea
\lefteqn{
\frac{1}{4}\sum_{s_{1},s_{2}}\sum_{\lambda_{1},\lambda_{2}}
|\langle W^{+}W^{-}|T|e^{+}e^{-}\rangle|^{2}} \nn \\
&=&
\frac{1}{4}\sum_{s_{1},s_{2}}
\biggl(\overline{v}\,T_{\mu'\nu'}u\biggr)^{*}
\biggl(-g^{\mu'\mu} + \frac{p_{1}^{\mu'}p_{1}^{\mu}}{M_{W}^{2}}\biggr)
\biggl(-g^{\nu'\nu} + \frac{p_{2}^{\nu'}p_{2}^{\nu}}{M_{W}^{2}}\biggr)
\biggl(\overline{v}\,T_{\mu\nu}u\biggr) \label{TT} \\
&=&
\frac{2\pi^{2}\alpha^{2}}{\sw^{4}}\sum_{i,j}M_{ij}\,,
\label{mij}
\eea
where $i,j = \gamma,Z,\nu$, and $\sw$ is the sine of the weak
mixing angle ($e = g\sw$, where $g$ is the SU(2)${}_{L}$ coupling).
In (\ref{mij}), the $M_{ij}$ are the various contributions
corresponding directly to the diagrams in Fig.~4.
For example, $M_{\gamma\gamma}$ is the contribution of the diagram
in Fig.~4(a), while $M_{Z\nu}$ is the $Z$--$\nu$ interference term.
These contributions were first
calculated almost twenty years ago by Alles, Boyer and Buras, 
and are given in Eqs.~(4.5)--(4.13) of their paper~\cite{allbowbur}.
The differential cross section is then given by
\be\label{dsigmaeeWW}
\frac{d\sigma(e^{+}e^{-}\rightarrow W^{+}W^{-})}{d\Omega} =
\frac{\alpha^{2}}{32\sw^{4}}\beta\frac{1}{s}
\sum_{i,j}M_{ij } \equiv
\sum_{i,j}\frac{d \sigma_{ij}}{d\Omega}\,.
\ee
Although the individual contributions $d\sigma_{ij}/d\Omega\propto M_{ij}$
in \eq{dsigmaeeWW} diverge at high energies,
the overall differential cross section
$d\sigma /d\Omega$ is well behaved as a result of extensive
cancellations among the various $d\sigma_{ij}/d\Omega$.
However, the bad high energy behaviour of the
$d\sigma_{ij}/d\Omega$ precludes the
use of the optical theorem (\ref{opth}) to interpret 
{\em individually} the components
$\{\sigma_{\gamma\gamma}, \sigma_{\gamma Z}, \sigma_{ZZ}\}$,
$\{\sigma_{\gamma\nu}, \sigma_{Z\nu}\}$
and $\{\sigma_{\nu\nu}\}$
of the tree level cross section for the process
$e^{+}e^{-}\ra W^{+}W^{-}$
in terms of the imaginary parts of renormalizable
one--loop self--energy--like, vertex--like and box--like
$W^{+}W^{-}$ contributions, respectively, to the
process $e^{+}e^{-} \ra e^{+}e^{-}$.

Here however, rather than decomposing the differential
cross section according to the
propagator structure of the Green's function $T^{\mu\nu}$,
as represented by the diagrams in Fig.~4,
we will decompose $d\sigma/d\Omega$  according to the propagator structure
occurring in the square of the modulus of the corresponding S--matrix element
$i\epsilon_{\mu}^{*}\epsilon_{\nu}^{*}\,\overline{v}\,T^{\mu\nu}u$.
The crucial observation is that these two decompositions are distinct.
This distinction is a fundamental consequence of the non--abelian structure
of the electroweak theory, and is encoded in the Ward identity obeyed by
$T^{\mu\nu}$. When the factors
of longitudinal four--momentum $p_{1\mu}$, $p_{2\nu}$
occurring in the $W^{-}$, $W^{+}$ polarizations sums in (\ref{TT})
are contracted with $T^{\mu\nu}$, this Ward identity
i) enforces the exact cancellation among the contributions from diagrams
with distinct $s$-- and $t$--channel propagator structure which 
individually diverge at high energy,
and ii) specifies the non--vanishing contributions which remain after this
cancellation and which are individually well--behaved at high energy.
It is these latter non--vanishing contributions
from the longitudinal factors in (\ref{TT}),
of purely non--abelian origin,\footnote{
In QED, the abelian gauge invariance of the theory is such that,
for any S--matrix element involving external photons,
the Ward identities ensure that the contribution due to a
longitudinal factor $p_{\mu}$ associated with the polarization vector
$\epsilon_{\mu}(p,\lambda)$ for an external photon vanishes identically.}
which are responsible for the 
distinction between the pole structure of the Green's function $T^{\mu\nu}$ 
and that occurring in the square of the modulus of the S--matrix element
$i\epsilon_{\mu}^{*}\epsilon_{\nu}^{*}\,\overline{v}\,T^{\mu\nu}u$.

The square of the modulus of the S--matrix element in (\ref{TT})
may thus be decomposed in terms of a new set of components $\hat{M}_{ij}$,
defined as before in terms of the propagator structures which occur,  
but only {\em after} the systematic
implementation of the Ward identity obeyed by $T^{\mu\nu}$
and triggered by the factors of longitudinal 
four--momentum $p_{1\mu}$, $p_{2\nu}$
occurring in the external $W^{-}$, $W^{+}$ polarization vectors.
The differential cross section is then given in terms of the
components $\hat{M}_{ij}$ by
\be\label{dhatsigmaeeWW}
\frac{d\sigma(e^{+}e^{-}\rightarrow W^{+}W^{-})}{d\Omega} =
\frac{\alpha^{2}}{32\sw^{4}}\beta\frac{1}{s}
\sum_{i,j}\hat{M}_{ij } \equiv
\sum_{i,j}\frac{d\hat{\sigma}_{ij}}{d\Omega}\,.
\ee
It is emphasized that the decomposition \eq{dhatsigmaeeWW} is 
uniquely specified by the Ward identity for $T^{\mu\nu}$.
As will be seen, the decomposition of the tree level cross section for 
$e^{+}e^{-} \ra W^{+}W^{-}$ 
in terms of the pole structure occurring in the modulus--squared
of the S--matrix element (\ref{T}), 
rather than that of the corresponding Green's function,
will enable the use of the optical theorem (\ref{opth}) 
to interpret {\em individually} the various components
$\hat{\sigma}_{ij}$ in terms
of the imaginary parts of renormalizable
one--loop self--energy--like, vertex--like and box--like
$W^{+}W^{-}$ contributions, respectively, to the
process $e^{+}e^{-} \ra e^{+}e^{-}$.

We begin by decomposing the
triple gauge vertex appearing in Figs.~4(a) and (b) as \cite{thooft}
\be
\Gamma_{\rho\mu\nu} =
\Gamma_{\rho\mu\nu}^{F} + \Gamma_{\rho\mu\nu}^{P}\, ,
\ee
where
\bea
\Gamma_{\rho\mu\nu}^{F}(q;-p_{1},-p_{2})&=&(p_{2}-p_{1})_{\rho}g_{\mu\nu}
- 2q_{\mu}g_{\rho\nu} + 2q_{\nu}g_{\rho\mu}\,, \\
\Gamma_{\rho\mu\nu}^{P}(q;-p_{1},-p_{2})&=&p_{1\mu}g_{\rho\nu} -
p_{2\nu}g_{\rho\mu}\,.
\eea
The component $\Gamma_{\rho\mu\nu}^{F}$
obeys a simple Ward identity involving
the difference of inverse gauge field propagators in the Feynman gauge:
$q^{\rho}\Gamma_{\rho\mu\nu}^{F}(q;-p_{1},-p_{2}) =
(p_{2}^{2} - p_{1}^{2})g_{\mu\nu}$.
The component $\Gamma_{\rho\mu\nu}^{P}$
vanishes when contracted into the polarization
vectors for the on--shell $W^{+}W^{-}$ pair:
$p_{1}^{\mu}\epsilon_{\mu}(p_{1},\lambda_{1}) =
p_{2}^{\nu}\epsilon_{\nu}(p_{2},\lambda_{2}) = 0$.
 
Choosing to work in the Feynman--'t Hooft gauge, the amputated Green's
function represented by the sum of the 
three diagrams in Fig.~4 may then be decomposed as \cite{papapil2,njw2} 
\be\label{TFP}
T_{\mu\nu} = T_{\mu\nu}^{F} + T_{\mu\nu}^{P}\,,
\ee
where 
\bea
T_{\mu\nu}^{F} 
&=&
% \biggl(ie^{2}\gamma^{\rho}\frac{1}{s}
% \,\,+\,\, ig^{2}\gamma^{\rho}(a-b\gamma_{5})\frac{1}{s-M_{Z}^{2}}
% \biggr)\Gamma_{\rho\mu\nu}^{F}(q;-p_{1},-p_{2}) \nn \\
%\label{TFmunu} & & -\frac{1}{8}ig^{2}
%\gamma_{\nu}(1-\gamma_{5})\frac{1}{\ks_{1} - \ps_{1}}
%\gamma_{\mu}(1-\gamma_{5})\,, \\ \nn \\
\label{TFmunu}
\biggl(ie^{2}\gamma^{\rho}\frac{1}{s}
+ ig^{2}\gamma^{\rho}(a-b\gamma_{5})\frac{1}{s-M_{Z}^{2}}
\biggr)\Gamma_{\rho\mu\nu}^{F}
-\frac{1}{8}ig^{2}
\gamma_{\nu}(1-\gamma_{5})\frac{1}{\ks_{1} - \ps_{1}}
\gamma_{\mu}(1-\gamma_{5})\,, \qquad \\
T_{\mu\nu}^{P} 
&=&
\biggl(ie^{2}\gamma^{\rho}\frac{1}{s}
+ ig^{2}\gamma^{\rho}(a-b\gamma_{5})\frac{1}{s-M_{Z}^{2}}
\biggr)\Gamma_{\rho\mu\nu}^{P}\,,
\label{TPmunu}
\eea
($a = \frac{1}{4} - \sw^{2}$, $b = \frac{1}{4}$).
The component $T_{\mu\nu}^{P}$
makes no contribution to the S--matrix element (\ref{T}).
Thus, in the expression (\ref{TT}), we only need to
consider the components $T_{\mu'\nu'}^{F}$, $T_{\mu\nu}^{F}$.\footnote{
It is worth remarking that the decomposition (\ref{TFP}) is not necessary,
but simply facilitates the calculation of the cross section.}
 
The next step is to evaluate the effect of the factors of longitudinal
four--momenta in \eq{TT}.
For a factor $p_{1}^{\mu}$, using the tree level Ward identities
\bea
p_{1}^{\mu}\Gamma_{\rho\mu\nu}^{F}(q;-p_{1},-p_{2}) &=&
- (s + p_{1}^{2}-p_{2}^{2})g_{\rho\nu}
+ q_{\rho}q_{\nu} + (p_{1}-p_{2})_{\rho}p_{2\nu}\, , \label{pGamma}\\
p_{1}^{\mu}\gamma_{\mu} &=& \ks_{1} - (\ks_{1} - \ps_{1})\, ,
\label{pgamma}
\eea

\begin{center}
\begin{picture}(400,310)(0,-20)
 
%\put(0,-10){\framebox(400,310){}}
 
\put( 20,220){\makebox(0,0)[c]{$p_{2}^{\nu}$}}
\put( 30,220){\vector(1,0){10}}
\put( 20,280){\makebox(0,0)[c]{$p_{1}^{\mu}$}}
\put( 30,280){\vector(1,0){10}}
 
\Photon( 60,220)( 75,250){4}{3}
\Photon( 60,280)( 75,250){4}{3}
\put( 90,265){\makebox(0,0)[c]{$\gamma$}}
\Photon( 75,250)(105,250){4}{3}
\Line(105,250)(120,220)
\Line(105,250)(120,280)
\put( 90,205){\makebox(0,0)[c]{(a)}}
 
\put( 20,120){\makebox(0,0)[c]{$p_{2}^{\nu}$}}
\put( 30,120){\vector(1,0){10}}
\put( 20,180){\makebox(0,0)[c]{$p_{1}^{\mu}$}}
\put( 30,180){\vector(1,0){10}}
 
\Photon( 60,120)( 75,150){4}{3}
\Photon( 60,180)( 75,150){4}{3}
\put( 90,165){\makebox(0,0)[c]{$Z$}}
\Photon( 75,150)(105,150){4}{3}
\Line(105,150)(120,120)
\Line(105,150)(120,180)
\put( 90,105){\makebox(0,0)[c]{(b)}}
 
\put(150,200){\makebox(0,0)[c]
{$ \left. \begin{array}{c}
\\ \\ \\ \\ \\ \\ \\ \\ \\ \\ \\ \\
\end{array} \right\}$}}

\put( 20, 20){\makebox(0,0)[c]{$p_{2}^{\nu}$}}
\put( 30, 20){\vector(1,0){10}}
\put( 20, 80){\makebox(0,0)[c]{$p_{1}^{\mu}$}}
\put( 30, 80){\vector(1,0){10}}
 
\Photon( 60, 20)( 90, 35){-4}{3}
\Photon( 60, 80)( 90, 65){4}{3}
\Line( 90, 35)( 90, 65)
\put(100, 50){\makebox(0,0)[c]{$\nu_{e}$}}
\Line( 90, 35)(120, 20)
\Line( 90, 65)(120, 80)
\put( 90,  5){\makebox(0,0)[c]{(c)}}
 
%%%%%%%%%
 
\put(200, 65){\makebox(0,0)[c]{\small pinch}}
\put(185, 50){\vector(1,0){30}}
 
\put(200,215){\makebox(0,0)[c]{\small pinch}}
\put(185,200){\vector(1,0){30}}
 
%%%%%%%%%%
 
\put(255,200){\makebox(0,0)[c]{\Large $+$}}
\Photon(280,170)(325,200){-4}{4.5}
\Photon(280,230)(325,200){4}{4.5}
\Line(325,200)(340,170)
\Line(325,200)(340,230)
\put(375,200){\makebox(0,0)[c]{\Large $+\,\,\,\cdots$}}
\put(315,155){\makebox(0,0)[c]{(d)}}
 
\put(255, 50){\makebox(0,0)[c]{\Large $-$}}
\Photon(280, 20)(325, 50){-4}{4.5}
\Photon(280, 80)(325, 50){4}{4.5}
\Line(325, 50)(340, 20)
\Line(325, 50)(340, 80)
\put(315,  5){\makebox(0,0)[c]{(e)}}
 
\end{picture}
\end{center}
 
\noindent
{\small Fig.~5.
The diagrammatic representation of the cancellation between
i) the component of the diagrams (a) and (b)
in which the $\gamma$ and $Z$ propagators
have been cancelled (pinched) by the action of the
factors of longitudinal four--momentum
$p_{1}^{\mu}$, $p_{2}^{\nu}$ appearing in the
$W^{-}$, $W^{+}$ polarization sums in \eq{TT},
and ii) the entire contribution of the diagram (c),
in which the $\nu_{e}$ propagator
has been cancelled (pinched) by the same factors.
The ellipsis in diagram (d) represents the remaining terms.}
 
\vspace{20pt}

\noindent
together with the fact that the external fields are on--shell (so that
$\overline{v}(k_{2},s_{2})\ks_{2} = \ks_{1} u(k_{1},s_{1}) = 0$ and
$p_{1}^{2} = p_{2}^{2} = M_{W}^{2}$), we obtain
\bea
p_{1}^{\mu}\,\overline{v}\,T_{\mu\nu}^{F}u 
&=&
g\sw\Bigl(-ie\,\overline{v}\gamma^{\rho}u\Bigr)
\frac{1}{s}(p_{2}-p_{1})_{\rho}p_{2\nu}  \nn \\ 
&+&
g\cw\Bigl(-i(g/\cw)\overline{v}\gamma^{\rho}(a - b\gamma_{5})u\Bigr)
\frac{1}{s-M_{Z}^{2}}
\Bigl(M_{Z}^{2}g_{\nu\rho} + (p_{2}-p_{1})_{\rho}p_{2\nu} \Bigr).
\label{pTF}
\eea
The simple dependence of this expression on the
$s$--channel $\gamma$ and $Z$ propagators
and the corresponding $ee\gamma$ and $eeZ$ couplings 
is apparent. The crucial feature of the Ward identity
Eq.\ (\ref{pTF}) is the {\em exact cancellation} which has occurred between
i) components of the
$\gamma$ and $Z$ contributions to $T_{\mu\nu}^{F}$
in Eq.\ (\ref{TFmunu}) in
which the $\gamma$ and $Z$ propagators have been cancelled (pinched) by the
term $sg_{\rho\nu}$ in Eq.~(\ref{pGamma}), and ii) the entire $\nu$
contribution to $T_{\mu\nu}^{F}$ in which the $\nu_{e}$ propagator 
has been cancelled (pinched) by the term
$\ks_{1}-\ps_{1}$ in Eq.~(\ref{pgamma}).
This cancellation is illustrated in Fig.~5.

Using the Ward identity 
Eq.~(\ref{pTF}) together with the similar expressions for the
various other longitudinal factors which occur in Eq.~(\ref{TT}), we can
isolate the contributions to the cross section according to the
$\gamma$, $Z$ and $\nu$ propagator structure
occurring in the square of the modulus of the S--matrix element,
as described above. In the centre of mass frame we obtain:
\bea
\label{Mgammagamma}
\hat{M}_{\gamma\gamma} &=&
\sw^{4}\Biggl\{12\beta^{2}\sin^{2}\theta - 64 \Biggr\}\,, \\
\label{MgammaZ}
\hat{M}_{\gamma Z} &=&
2\sw^{2}a\frac{s}{(s-M_{Z}^{2})}
\Biggl\{ \biggl(12 - 2\frac{1}{\cw^{2}} \biggr)
\beta^{2}\sin^{2}\theta -64 \Biggl\}\,, \\
\label{MZZ}
\hat{M}_{ZZ} &=& (a^{2}+b^{2})\frac{s^{2}}{(s-M_{Z}^{2})^{2}}
\Biggl\{ \biggl(12 - 4\frac{1}{\cw^{2}} + \frac{1}{\cw^{4}}
\biggr)\beta^{2}\sin^{2}\theta -64 +
16\frac{1}{\cw^{4}}\frac{M_{W}^{2}}{s} \Biggr\}\,; \\
& & \phantom{\large A} \nn \\
\label{Mgammanu}
\hat{M}_{\gamma\nu} &=& 2\sw^{2}\frac{s}{t}
\Biggl\{ \beta^{2}\sin^{2}\theta - 4(1 -
\beta\cos\theta)\Biggr\}\,, \\
\label{MZnu}
\hat{M}_{Z\nu} &=& 2(a+b)\frac{s}{(s-M_{Z}^{2})}\frac{s}{t}
\Biggl\{ \beta^{2}\sin^{2}\theta
- 4(1 - \beta\cos\theta)\Biggr\}\,; \\
& & \phantom{\large A} \nn \\
\label{Mnunu}
\hat{M}_{\nu\nu} &=&
\frac{1}{2}\frac{s^{2}}{t^{2}}\beta^{2}\sin^{2}\theta\,.
\eea
The above expressions have been grouped into self--energy--like,
vertex--like and box--like contributions
according to the presence of zero, one or two internal
$\nu_{e}$ propagators, respectively.
In each case, an overall factor $\vartheta(s - 4M_{W}^{2})$
has been omitted for brevity.
 
Carrying out the angular integrations, the corresponding contributions
to the total cross section are given by
\bea
\label{sigmagammagamma}
\hat{\sigma}_{\gamma\gamma} &\!=\!& 2\pi\alpha^{2}
\frac{1}{s}\beta\Biggl\{ -\frac{7}{2} - \frac{2}{y}\Biggr\}\,, \\
\label{sigmagammaZ}
\hat{\sigma}_{\gamma Z}
&\!=\!& 4\pi\alpha^{2}\frac{a}{\sw^{2}\cw^{2}}
\frac{1}{(s- M_{Z}^{2})}\beta
\Biggl\{ -\frac{43}{12} - \frac{5}{3y}
-s_{\rm w}^{2}\biggl(\! -\frac{7}{2} - \frac{2}{y} \biggr)\Biggr\}\,, \\
\label{sigmaZZ}
\hat{\sigma}_{ZZ} &\!=\!& 2\pi\alpha^{2}\frac{(a^{2} + b^{2})}{\sw^{4}\cw^{4}}
\frac{s}{(s-M_{Z}^{2})^{2}}\beta
\Biggl\{ -\frac{29}{8} - \frac{1}{2y} -2s_{\rm w}^{2}
\biggl(\!-\frac{43}{12} - \frac{5}{3y}\biggr) +
s_{\rm w}^{4}\biggl(\!-\frac{7}{2} - \frac{2}{y}\biggr)
\Biggr\}\,;\,\,\,\,\,\,\,\,\,\,\,\, \\ 
& & \phantom{\large A} \nn \\ 
\label{sigmagammanu}
\hat{\sigma}_{\gamma\nu} &\!=\!& 4\pi\alpha^{2}\frac{1}{\sw^{2}}
\frac{1}{s}\beta\Biggl\{
\biggl(\frac{1}{y} + \frac{1}{2y^{2}}\biggr)
\frac{L}{\beta} + \frac{3}{8} + \frac{1}{4y}
\Biggr\}\,, \\
\label{sigmaZnu}
\hat{\sigma}_{Z\nu} &\!=\!& 4\pi\alpha^{2}\frac{(a+b)}{\sw^{4}}
\frac{1}{(s-M_{Z}^{2})}\beta
\Biggl\{\biggl(\frac{1}{y} + \frac{1}{2y^{2}}\biggr)
\frac{L}{\beta} + \frac{3}{8} + \frac{1}{4y}
\Biggr\}\,; \\ 
& & \phantom{\large A} \nn \\ 
\label{sigmanunu}
\hat{\sigma}_{\nu\nu} &\!=\!& 2\pi\alpha^{2}\frac{1}{\sw^{4}}
\frac{1}{s}\beta
\Biggl\{ \biggl(\frac{1}{4} - \frac{1}{2y}\biggr)\frac{L}{\beta}
-\frac{1}{4}\Biggr\}\,.
\eea
Here $y \equiv s/M_{W}^{2}$, $L \equiv \ln[(1+\beta)/(1-\beta)]$
and again the factor $\vartheta(s - 4M_{W}^{2})$ is omitted.

\pagebreak

\centerline{\epsfxsize = 350pt \epsfbox{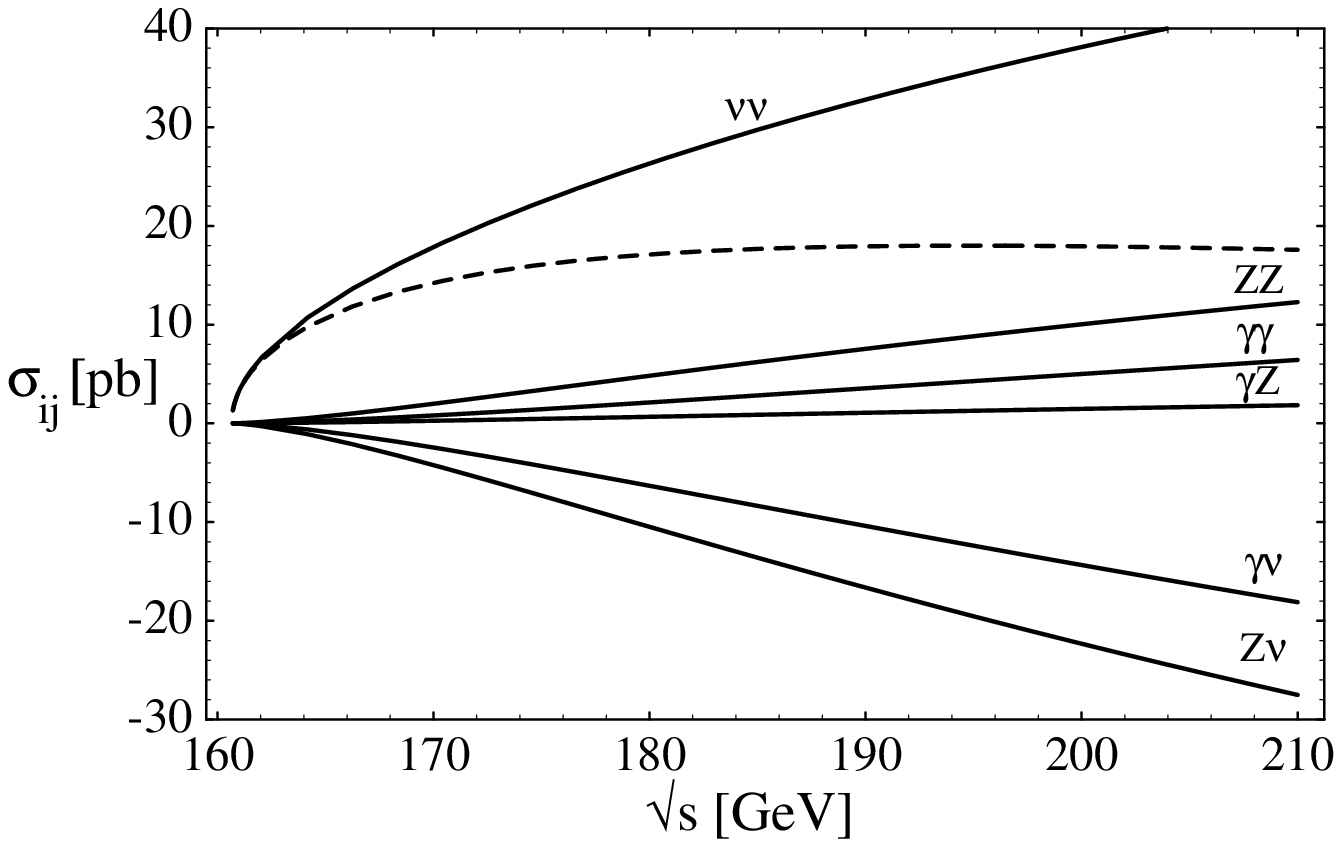}}

\vspace{10pt}

\noindent
{\small Fig.~6(a).
The components $\sigma_{ij}$, $i,j = \gamma,Z,\nu$,
of the tree level cross section
$\sigma(e^{+}e^{-} \ra W^{+}W^{-})$, corresponding to the 
decomposition directly in terms of the diagrams in Fig.~4, 
taken from Ref.~\cite{allbowbur}. The dashed line is the 
full cross section, given by the sum of the components.}

\vspace{20pt}
 
\centerline{\epsfxsize = 350pt \epsfbox{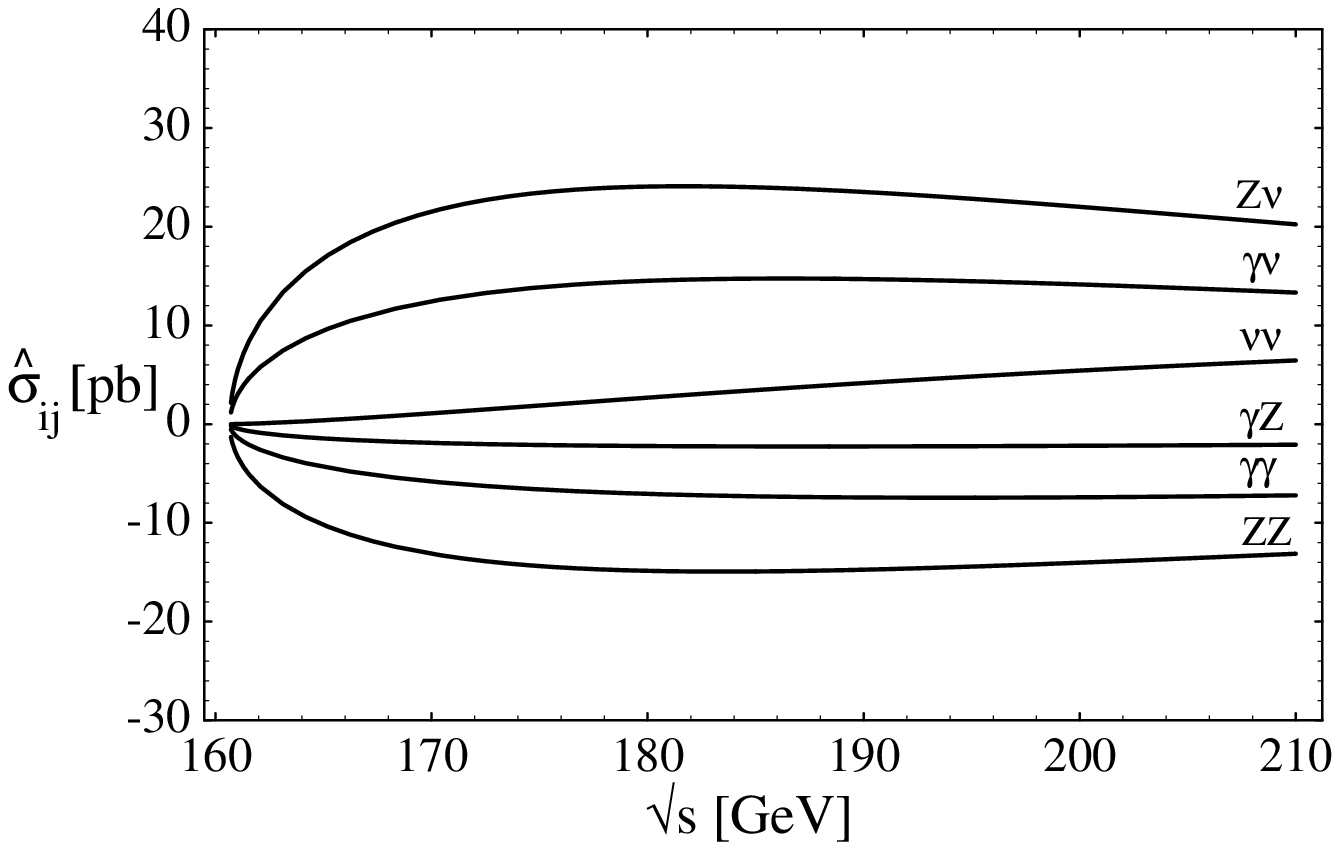}}

\vspace{10pt}

\noindent
{\small Fig.~6(b).
The components $\hat{\sigma}_{ij}$,
$i,j = \gamma,Z,\nu$, of the tree level cross section
$\sigma(e^{+}e^{-} \ra W^{+}W^{-})$, corresponding to the
decomposition according to the propagator structure occurring
in the square of the modulus of the S--matrix element, and
given in Eqs.~(\ref{sigmagammagamma})--(\ref{sigmanunu}).}

\vspace{20pt}
 
The cross section components $\hat{\sigma}_{ij}$ given in
Eqs.~(\ref{sigmagammagamma})--(\ref{sigmanunu})
are shown plotted in Fig.~6(b). These plots are to be compared with 
the plots of the components $\sigma_{ij}$ obtained by Alles, Boyer and Buras
\cite{allbowbur} shown in Fig.~6(a).

\pagebreak

Several remarks are in order:
\begin{itemize}
\item It is easily verified that these expressions Eqs.\
(\ref{Mgammagamma})--(\ref{Mnunu}) and
(\ref{sigmagammagamma})--(\ref{sigmanunu}) combine to reproduce,
respectively,
the correct differential and total cross sections given in \cite{allbowbur}.
\item At asymptotic $s$, each of the contributions
$\hat{M}_{ij} \sim {\rm const}$. This is in contrast to the conventional
decomposition, in which, as a result of the longitudinal factors in the
$W^{+}$, $W^{-}$ polarization sums, each of the $M_{ij} \sim s^{2}$ so
that individually they violate unitarity; it is only when summed that, as a
result of cancellations, $\sum_{ij}M_{ij} \sim {\rm const.}$ and unitarity is
restored \cite{dirgen,cornlevtikt}. Here, by systematically implementing
on the r.h.s.\ of \eq{TT}
the Ward identity \eq{pTF} triggered by these longitudinal factors,
this cancellation has occurred at the level
of the individual components $\hat{M}_{ij}$ of the differential cross
section rather than the overall differential cross section itself.
 
\item The three self--energy--like cross section contributions
$\hat{\sigma}_{\gamma\gamma}$,
$\hat{\sigma}_{\gamma Z}$ and
$\hat{\sigma}_{ZZ}$ are each negative.
This is directly related to the fact that the $W^{\pm}$ gauge boson
contributions to the one--loop electroweak $\beta$--functions
are negative. In particular,
the coefficient $-\frac{7}{2} = -\frac{11}{6}\!\times\! 2 + \frac{1}{6}$
in the expression Eq.~(\ref{sigmagammagamma}) for
$\hat{\sigma}_{\gamma\gamma}$ is precisely
the one--loop contribution of the massive $W^{\pm}$ gauge bosons to the
$\beta$--function for the electromagnetic coupling $\alpha$.
Similarly, writing  $J_{\gamma}J_{Z}
= (J_{3} + \frac{1}{2}J_{Y})J_{3} -\sw^{2}J_{\gamma}J_{\gamma}$
where $J_{\gamma}$, $J_{Z}$, $J_{3}$ and $J_{Y}$
are the currents which couple
to the $\gamma$, $Z$, $W_{3}$ and $B$ gauge fields respectively
($W_{3}$ is the neutral component of the SU$(2)_{L}$ triplet, while
$B$ is the U$(1)_{Y}$ singlet), the coefficient
$-\frac{43}{12} = -\frac{11}{6}\!\times\!2 + \frac{1}{12}$
in the expression Eq.~(\ref{sigmagammaZ}) for $\hat{\sigma}_{\gamma Z}$
is the one--loop contribution of
the massive $W^{\pm}$ gauge bosons to the
$\beta$--function for the SU$(2)_{L}$ coupling $g^{2}/4\pi$.
In each case, the term $-\frac{11}{6}\!\times\!2$ is the pure gauge
field contribution, while the terms $+\frac{1}{6}$ and $+\frac{1}{12}$
respectively are the contributions of the complex doublet of
scalars involved in the spontaneous symmetry breaking 
(cf.\ Sec.~6). The appearance of the $W^{\pm}$ gauge boson
contributions to the electroweak $\beta$--functions
in the $e^{+}e^{-}\ra W^{+}W^{-}$ cross section components
$\hat{\sigma}_{\gamma\gamma}$ and $\hat{\sigma}_{\gamma Z}$
is exactly analogous to the appearance
of the $f^{\pm}$ fermion contribution to the QED $\beta$--function in the
$e^{+}e^{-}\ra f^{+}f^{-}$ cross section Eq.~(\ref{sigmaff2}).
\end{itemize}
 
%%%%%%%%%%%%%%%%%%%%%%%%%%%%%%%%
 
\subsection{$e^{+}e^{-} \rightarrow ZH$}
 
We next consider the process
$e^{-}(k_{1},s_{1})e^{+}(k_{2},s_{2})
\rightarrow Z(p_{1},\lambda_{1})H(p_{2})$
at tree level, where $H$ is the Higgs scalar.
Neglecting the fermion mass, the
only diagram which contributes to this process is that shown in Fig.~7.
 
A straightforward calculation gives for the cross section
\be\label{sigmaZH}
\sigma(e^{+}e^{-}\!\rightarrow\! ZH)
=
2\pi\alpha^{2}\frac{(a^{2}\!+\!b^{2})}{\sw^{4}\cw^{4}}
\frac{s}{(s-M_{Z}^{2})^{2}}
\lambda^{\frac{1}{2}}(s,M_{Z},M_{H})
\Biggl\{ \frac{1}{24} + \frac{5M_{Z}^{2}}{12s} - \frac{M_{H}^{2}}{12s}
+ \frac{(M_{Z}^{2}\!-\!M_{H}^{2})^{2}}{24s^{2}} \Biggr\}
\ee
where $M_{H}$ is the Higgs mass and
$\lambda(s,M_{i},M_{j}) = (1 - (M_{i}+M_{j})^{2}/s)(1 - (M_{i}-M_{j})^{2}/s)$.

\begin{center}
\begin{picture}(400,130)(0,20)
 
%\put(0, 20){\framebox(400,180){}}
 
\put(164, 58){\makebox(0,0)[c]{\small $e^{+}(k_{2},s_{2})$}}
\put(164,144){\makebox(0,0)[c]{\small $e^{-}(k_{1},s_{1})$}}
\ArrowLine(170, 70)(185,100)
\ArrowLine(170,130)(185,100)
\put(203,115){\makebox(0,0)[c]{\small $Z$}}
\Photon(185,100)(215,100){4}{3}
\DashArrowLine(215,100)(230, 70){6}
\Photon(215,100)(230,130){4}{3}
\ArrowLine(221,115)(226,115)
\put(241, 58){\makebox(0,0)[c]{\small $H(p_{2})$}}
\put(241,144){\makebox(0,0)[c]{\small $Z(p_{1},\lambda_{1})$}}

\end{picture}
\end{center}
 
\noindent
{\small Fig.~7.
The single diagram which contributes to the process
$e^{+}e^{-} \rightarrow ZH$ at tree level for the 
case of massless fermions.}
 
\vspace{20pt}

Two remarks are in order:
\begin{itemize}
\item The calculation of the cross section Eq.~(\ref{sigmaZH}) involves
only a single diagram, so that, in contrast to the process
$e^{+}e^{-}\rightarrow W^{+}W^{-}$, there is no cancellation
to account for involving factors of external gauge field
longitudinal four--momentum.
\item In the high energy limit $s/M_{Z}^{2}\ra\infty$,
the sum of the $e^{+}e^{-}\ra W^{+}W^{-}$
self--energy--like cross section component
$\hat{\sigma}_{ZZ}$ Eq.~(\ref{sigmaZZ}) and the
$e^{+}e^{-}\rightarrow ZH$ cross section Eq.~(\ref{sigmaZH}) is given by
\bea
\lefteqn{
\hat{\sigma}_{ZZ}(e^{+}e^{-}\rightarrow W^{+}W^{-})
\,\,+\,\,
\sigma(e^{+}e^{-}\rightarrow ZH) \,\,\,= } \nn \\
\label{JZJZ}
& & \hspace{20pt}
2\pi\alpha^{2}\frac{(a^{2}+b^{2})}{\sw^{4}\cw^{4}}\frac{1}{s}
\Biggl\{ -\frac{43}{12} -2\sw^{2}\biggl(-\frac{43}{12}\biggr)
+ \sw^{4}\biggl(-\frac{7}{2} \biggr) +
\cO\biggl(
\frac{M_{Z}^{2}}{s},
\frac{M_{H}^{2}}{s}\biggr)\Biggl\}.\,\,\,\,
\eea
Writing
$J_{Z}J_{Z} = J_{3}J_{3} -2\sw^{2}(J_{3} + \frac{1}{2}J_{Y})J_{3} +
\sw^{4}J_{\gamma}J_{\gamma}$, the coefficient
$-\frac{43}{12} = -\frac{29}{8} + \frac{1}{24} =
-\frac{11}{6}\!\times\! 2 + \frac{1}{12}$ in \eq{JZJZ}
is again precisely the one--loop contribution of the massive electroweak
gauge bosons to the $\beta$--function
for the SU$(2)_{L}$ coupling $g^{2}/4\pi$ (cf.\ Sec.~6). 
\end{itemize}
 
%%%%%%%%%%%%%%%%%%%%%%%%%%%%%%%%%%%%%%%%%%%%%%%%%
 
\setcounter{equation}{0}
 
\section{The Electroweak Gauge Boson Self-Energies}
 
We now turn to the gauge boson self-energies.
In general, the transverse self--energy
function $\Sigma_{ij}$ for gauge bosons $i$, $j$
requires two subtractions, for mass and field renormalization\footnote{
For a thorough review of radiative corrections in electroweak theory,
where many references may be found, see e.g.\ Ref.~\cite{hollik}}:
with subtraction point $s_{ij}$,
the renormalized self--energy $\Sigma_{R,ij}$ is given by
\bea\label{sigmaR}
\Sigma_{R,ij}(q^2)
&=&
\Sigma_{ij}(q^2) - \Sigma_{ij}(s_{ij})
- (q^{2} - s_{ij})
\left.\frac{\partial\Sigma_{ij}(q^2)}{\partial q^{2}}\right|_{s_{ij}} \\
&=&
\label{sigmadisprel}
(q^2-s_{ij})^{2}\int_{0}^{\infty}
ds \frac{1}{(s- q^{2})(s- s_{ij})^{2}}
\frac{1}{\pi}\,\im\Sigma_{ij}(s)\,.
\eea
In the case where $s_{ij} = 0$
and $\Sigma_{ij}(0) = 0$, writing $\Sigma_{ij}(q^2) = q^{2}\Pi_{ij}(q^2)$,
the twice--subtracted dispersion relation Eq.~(\ref{sigmadisprel})
reduces to a once--subtracted expression identical to Eq.~(\ref{qeddisprel})
for the QED vacuum polarization renormalized in the on--shell scheme.
 
In the conventional perturbation theory approach, the gauge boson
plus associated would--be Goldstone boson and ghost
contributions to the electroweak self--energies
$\Sigma_{ij}$ are gauge--dependent
(for expressions in the class of renormalizable gauges, 
see e.g.\ Refs.~\cite{passveltdegsir2}).
Here however, rather than considering the contributions to
$\Sigma_{ij}$ from one--loop perturbation theory diagrams in a given gauge,
we will follow 
the dispersive procedure described in Sec.~2 for the case of QED to 
{\em define} one--loop electroweak self--energy--like functions
$\hat{\Sigma}_{ij}$ directly from the tree level
cross sections for physical processes.
In particular, having decomposed the tree level cross section
for $e^{+}e^{-}\rightarrow W^{+}W^{-}$ in terms of the pole
structure occurring in
the square of the modulus of the corresponding S--matrix element,
we define the imaginary parts of the $W^{+}W^{-}$
contributions to the $\hat{\Sigma}_{ij}$ via
\bea\label{imsigmaWWgammagamma}
e^{2} \frac{1}{s^{2}}\,
\im\hat{\Sigma}_{\gamma\gamma}^{\scriptscriptstyle (WW)}(s)
&=&
\hat{\sigma}_{\gamma\gamma}(e^{+}e^{-}\ra W^{+}W^{-})\,,
\\
\label{imsigmaWWgammaZ}
2e^{2}\frac{a}{\sw\cw}\,\frac{1}{s(s-M_{Z}^{2})}\,
\im\hat{\Sigma}_{\gamma Z}^{\scriptscriptstyle (WW)}(s)
&=&
\hat{\sigma}_{\gamma Z}(e^{+}e^{-}\ra W^{+}W^{-})\,,
\\
\label{imsigmaWWZZ}
e^{2}\frac{(a^{2} + b^{2})}{\sw^{2}\cw^{2}}\,\frac{1}{(s-M_{Z}^{2})^{2}}\,
\im\hat{\Sigma}_{ZZ}^{\scriptscriptstyle (WW)}(s)
&=&
\hat{\sigma}_{ZZ}(e^{+}e^{-}\ra W^{+}W^{-})\,.
\eea
Similarly, we define the imaginary part of the $ZH$ contribution
to $\hat{\Sigma}_{ZZ}$ via
\be\label{imsigmaZHZZ}
e^{2}\frac{(a^{2} + b^{2})}{\sw^{2}\cw^{2}}\,\frac{1}{(s-M_{Z}^{2})^{2}}\,
\im\hat{\Sigma}_{ZZ}^{\scriptscriptstyle (ZH)}(s)
\,\,\,=\,\,\,
\sigma(e^{+}e^{-}\ra ZH)\,.
\ee
These definitions are by direct analogy with the QED relation
Eq.~(\ref{sigmaff1}) (with $m_{e} = 0$) between the imaginary part 
of the fermion contribution to the one--loop photon
self--energy $\Sigma(q^2) = q^{2}\Pi(q^2)$ and the tree level
cross section for $e^{+}e^{-} \rightarrow f^{+}f^{-}$:
in each case there is a factor consisting of the product of propagators
$(q^{2} - M_{i}^{2})^{-1}(q^{2} - M_{j}^{2})^{-1}$ multiplying the
couplings of the corresponding gauge fields $i$, $j$ to the $e^{+}e^{-}$ pair.
Given the contribution of a particular pair of fields to
$\im\hat{\Sigma}_{ij}$ together with an appropriate
choice for the subtraction point $s_{ij}$, the contribution
of the fields to the renormalized
self--energy function $\hat{\Sigma}_{R,ij}$ can then be reconstructed using
the dispersion relation Eq.~(\ref{sigmadisprel}).

It is emphasized that the analogy with the QED relation \eq{sigmaff1}
uniquely specifies the $W^{+}W^{-}$ contributions to the 
one--loop self--energies
in terms of the self--energy--like components of the tree level 
$e^{+}e^{-} \ra W^{+}W^{-}$ 
cross section, which are in turn uniquely specified by the 
systematic and exhaustive implementation of the Ward identity for
the corresponding tree level amputated Green's function,
as described in Sec.~3.1.
In particular, the renormalized self--energies $\hat{\Sigma}_{ij}$ 
defined in this way in terms of the cross sections for physical
processes are clearly gauge--independent.
Furthermore, we will show how,
by defining the electroweak self--energies by direct analogy with
the properties (ii) and (iii) of the QED self--energy
described in Sec.~2,
not only does the basic QED property (i) of gauge independence extend
to the electroweak theory (to one--loop level), but
also each of the further properties (iv)--(vii).

We now choose the on--shell renormalization scheme.
The subtraction points are then
$s_{\gamma\gamma} = s_{\gamma Z} = 0$ and $s_{ZZ} = M_{Z}^{2}$.
Using the definitions
Eqs.~(\ref{imsigmaWWgammagamma})--(\ref{imsigmaWWZZ}), the imaginary
parts of the $W^{+}W^{-}$ contributions to
$\hat{\Sigma}_{\gamma\gamma}$, $\hat{\Sigma}_{\gamma Z}$ and
$\hat{\Sigma}_{ZZ}$ may be read off from the
$e^{+}e^{-}\ra W^{+}W^{-}$ cross section contributions
$\hat{\sigma}_{\gamma\gamma}$, $\hat{\sigma}_{\gamma Z}$ and
$\hat{\sigma}_{ZZ}$, respectively, given in
Eqs.~(\ref{sigmagammagamma})--(\ref{sigmaZZ}). Inserting 
these expressions  in the dispersion 
relation Eq.~(\ref{sigmadisprel}) and 
carrying out the integrations, we obtain:
\bea
\lefteqn{
\hat{\Sigma}_{R,\gamma\gamma}^{\scriptscriptstyle (WW)}(q^2)
\,\,\,=\,\,\,
\frac{\alpha}{\pi}\,q^{2}\,
\Biggl\{ \biggl(\frac{7}{2} + \frac{2M_{W}^{2}}{q^2}\biggr)B(q^2)
+ \frac{1}{6} \Biggr\} \,,} \,\,\,\, \label{SigmaRggWW} \\
\lefteqn{
\hat{\Sigma}_{R,\gamma Z}^{\scriptscriptstyle (WW)}(q^2)
\,\,\,=\,\,\,
\frac{\alpha}{\pi}\frac{1}{\sw\cw}\,q^{2}\,
\Biggl\{
\biggl[\biggl(\frac{43}{12} + \frac{5M_{W}^{2}}{3q^{2}}\biggr)
-\sw^{2}\biggl(\frac{7}{2} + \frac{2M_{W}^{2}}{q^2}\biggr)\biggr]B(q^2)
+ \frac{5}{36} -\sw^{2}\frac{1}{6} \Biggr\} \,,}\,\,\,\,\label{SigmaRgZWW}\\
\lefteqn{
\hat{\Sigma}_{R,ZZ}^{\scriptscriptstyle (WW)}(q^2)
\,\,\,=\,\,\,
\frac{\alpha}{\pi}\frac{1}{\sw^{2}\cw^{2}}
\,\Bigl(q^{2}-M_{Z}^{2}\Bigr) \times } \nn \,\,\,\,\\
& &
\Biggl\{
\biggl[\biggl(\frac{29}{8} + \frac{M_{W}^{2}}{2q^2}\biggr)
-2\sw^{2}\biggl(\frac{43}{12} + \frac{5M_{W}^{2}}{3q^2}\biggr)
+\sw^{4}\biggl(\frac{7}{2} + \frac{2M_{W}^{2}}{q^2}\biggr)\biggr]
q^{2}\biggl(\frac{B(q^2)-B(M_{Z}^{2})}{q^2-M_{Z}^{2}}\biggr) \nn \\
& &
-\biggl[\biggl(\frac{29}{8} + \frac{1}{2}\cw^{2}\biggr)
-2\sw^{2}\biggl(\frac{43}{12} + \frac{5}{3}\cw^{2}\biggr)
+\sw^{4}\biggl(\frac{7}{2} + 2\cw^{2}\biggr)\biggr]
M_{Z}^{2}\left.\frac{dB(q^2)}{dq^{2}}\right|_{M_{Z}^{2}}
\Biggr\}\,, \label{SigmaRZZWW}
\eea
where the function $B(q^2)$ is given by
\be\label{B}
B(q^2)
=
\frac{\beta}{2}\ln \biggl(\frac{\beta+1}{\beta-1}\biggr) -1\,, \qquad
\beta = \sqrt{1-\frac{4M_{W}^{2}}{q^2}}.
\ee
Thus,
\bea
\im B(q^{2})
&=&
-\frac{\pi}{2}\beta \,\vartheta(q^{2} - 4M_{W}^{2})\,, \\
q^{2}\,\frac{dB(q^2)}{dq^{2}}
&=&
\frac{1}{\beta^{2}}\biggl( \frac{2M_{W}^{2}}{q^{2}} B(q^{2}) +
\frac{1}{2}\biggr)\,.
\eea
For asymptotic $q^2$,
\be
B(q^2)
=
\left\{
\begin{array}{ll}
{\displaystyle
-\frac{1}{12}\frac{q^2}{M_{W}^{2}}
-\frac{1}{120}\frac{q^{4}}{M_{W}^{4}}
+{\cal O}\biggl(\frac{q^{6}}{M_{W}^{6}}\biggr)}
& \hspace{30pt}{\displaystyle
\phantom{-}q^2/M_{W}^{2} \rightarrow 0} \\
{\displaystyle
\frac{1}{2}\ln\biggl(-\frac{q^2}{M_{W}^{2}}\biggr) - 1
+{\cal O}\biggl(\frac{M_{W}^{2}}{q^2}\biggr)}
& \hspace{30pt}{\displaystyle
-q^2/M_{W}^{2} \rightarrow \infty}\,.
\end{array}
\right.
\ee
 
Similarly, using the definition \eq{imsigmaZHZZ}, the imaginary
part of the $ZH$ contribution to $\hat{\Sigma}_{R,ZZ}$ may be read
off from the cross section  for $e^{+}e^{-}\ra ZH$
given in \eq{sigmaZH}.
The function $\hat{\Sigma}_{R,ZZ}^{\scriptscriptstyle (ZH)}(q^2)$
may then be obtained from the dispersion relation \eq{sigmadisprel}.
The resulting expression is given in App.~A.
 
The expressions Eqs.~(\ref{SigmaRggWW})--(\ref{SigmaRZZWW}) for
$\hat{\Sigma}_{R,\gamma\gamma}^{\scriptscriptstyle (WW)}$,
$\hat{\Sigma}_{R,\gamma Z}^{\scriptscriptstyle (WW)}$  and
$\hat{\Sigma}_{R,ZZ}^{\scriptscriptstyle (WW)}$ and
\eq{SigmaRZZZH} for $\hat{\Sigma}_{R,ZZ}^{\scriptscriptstyle (ZH)}$
are {\em identical} to the corresponding renormalized
gauge--independent self--energy contributions
obtained in the pinch technique \cite{degsir}.
In the usual pinch technique approach, these self--energy
contributions were
obtained via the rearrangement of conventional one--loop
perturbation theory diagrams. Here, we have obtained
exactly the same results directly from components of the
tree level cross sections for physical processes.
 
An exactly similar procedure to that described above can be carried
out for the charged current interactions in order to define
the gauge boson contributions to the
renormalized one--loop $W$ self--energy $\hat{\Sigma}_{R,WW}$.
The relation between the imaginary part of $\hat{\Sigma}_{R,WW}$
and the tree level cross sections for the corresponding
physical processes is specified by the optical theorem
for the case of forward scattering in the process
$e^{+}\nu_{e}\ra e^{+}\nu_{e}$ :
\be\label{opthcc}
\im \langle e^{+}\nu_{e}|T|e^{+}\nu_{e} \rangle =
\frac{1}{2}\sum_{i}\int d\Gamma_{i}\,|\langle e^{+}\nu_{e}|T|i\rangle |^{2}.
\ee
The on--shell states involving the electroweak gauge bosons $W$ and $Z$
which contribute to the r.h.s.\ of \eq{opthcc} at lowest order are
$|W^{+}Z\rangle$, $|W^{+}\gamma\rangle$ and $|W^{+}H\rangle$,
corresponding to the tree level physical processes
$e^{+}\nu_{e}\ra W^{+}Z$,
$e^{+}\nu_{e}\ra W^{+}\gamma$ and
$e^{+}\nu_{e}\ra W^{+}H$.

For the processes
$e^{+}\nu_{e}\ra W^{+}Z$  and $e^{+}\nu_{e}\ra W^{+}\gamma$,
an exactly similar cancellation occurs among
contributions to the cross section involving factors of external
gauge field longitudinal four--momentum as occurred in the
process $e^{+}e^{-}\ra W^{+}W^{-}$ analysed in Sec.~3.1.
In each case, the square of the modulus of the
S--matrix element may be decomposed according to the pole structure
which occurs {\em after} the systematic implementation of the
Ward identity obeyed by the corresponding Green's function.
In particular, the self--energy--like components
$\hat{\sigma}_{WW}(e^{+}\nu_{e}\ra W^{+}Z)$ and
$\hat{\sigma}_{WW}(e^{+}\nu_{e}\ra W^{+}\gamma)$ of the
two cross sections
are the contributions after cancellations which
involve a pair of $s$--channel $W$ propagators $(s - M_{W}^{2})^{-1}$.
The imaginary parts of the $WZ$ and $W\gamma$ contributions
to the $W$ self--energy $\hat{\Sigma}_{WW}$ are then defined via
($m_{e} = 0$)
\bea\label{imsigmaWZWW}
e^{2}\frac{1}{4\sw^{2}}\,\frac{1}{(s-M_{W}^{2})^{2}}\,
\im\hat{\Sigma}_{WW}^{\scriptscriptstyle (WZ)}(s)
&=&
\hat{\sigma}_{WW}(e^{+}\nu_{e}\ra W^{+}Z)\,, \\
\label{imsigmaWgammaWW}
e^{2}\frac{1}{4\sw^{2}}\,\frac{1}{(s-M_{W}^{2})^{2}}\,
\im\hat{\Sigma}_{WW}^{\scriptscriptstyle (W\gamma)}(s)
&=&
\hat{\sigma}_{WW}(e^{+}\nu_{e}\ra W^{+}\gamma)\,.
\eea
For the process $e^{+}\nu_{e}\ra W^{+}H$, involving only a single
diagram at tree level, there is no cancellation to account for. The
imaginary part of the $WH$ contribution to
$\hat{\Sigma}_{WW}$ is defined via
\be\label{imsigmaWHWW}
e^{2}\frac{1}{4\sw^{2}}\,\frac{1}{(s-M_{W}^{2})^{2}}\,
\im\hat{\Sigma}_{WW}^{\scriptscriptstyle (WH)}(s)
\,\,\,=\,\,\,
\sigma (e^{+}\nu_{e}\ra W^{+}H)\,.
\ee

In the on--shell renormalization scheme,
the subtraction point $s_{WW} = M_{W}^{2}$.
The $WZ$, $W\gamma$ and $WH$ contributions to the
renormalized one--loop $W$ self--energy
$\hat{\Sigma}_{WW}$ can then be reconstructed from the
expressions for
$\hat{\sigma}_{WW}(e^{+}\nu_{e}\ra W^{+}Z)$,
$\hat{\sigma}_{WW}(e^{+}\nu_{e}\ra W^{+}\gamma)$
and $\sigma (e^{+}\nu_{e}\ra W^{+}H)$, respectively,
using the dispersion relation \eq{sigmadisprel}
in conjunction with the above definitions.
The resulting expressions for the functions
$\hat{\Sigma}_{WW}^{\scriptscriptstyle (WZ)}$,
$\hat{\Sigma}_{WW}^{\scriptscriptstyle (W\gamma)}$ and
$\hat{\Sigma}_{WW}^{\scriptscriptstyle (WH)}$
are given in App.~A.
These functions are again {\em identical} to the corresponding
renormalized gauge--independent self--energy contributions
obtained in the pinch technique \cite{degsir}.

It is important to point out that the PT one--loop gauge boson
self--energy functions are universal~\cite{njw3}.
This is a direct result of the tree level Ward identities of the
given theory \cite{papapil2,njw2}.
In the approach followed here, this universality corresponds
to the fact that the imaginary parts of the
one--loop contributions of fields $C$, $D$ to the PT self--energies
may be obtained from the self--energy--like components
of the tree level cross section $\sigma(AB \ra CD)$
for the interaction of {\em any} pair
of fields $A$, $B$ to which the gauge bosons couple at tree level.
For example, the $W^{+}W^{-}$ contributions to
$\hat{\Sigma}_{\gamma\gamma}$,
$\hat{\Sigma}_{\gamma Z}$ and $\hat{\Sigma}_{ZZ}$
may just as well be obtained from the self--energy--like components
of the cross section $\sigma(W^{+}W^{-} \ra W^{+}W^{-})$
as from the cross section 
$\sigma(e^{+}e^{-} \ra W^{+}W^{-})$ considered here.

%%%%%%%%%%%%%%%%%%%%%%%%%%%%%%%%%%%%%%%%%%%%%%%%%%
 
\setcounter{equation}{0}
 
\section{The Electroweak Effective Charges}
 
We next turn to the electroweak effective charges.
Beyond tree level, the $\gamma$--$Z$ mixing induced by the
$\hat{\Sigma}_{\gamma Z}$ self--energy requires the neutral current
sector to be re--diagonalized.
In Refs.~\cite{papapil1,njw2},
it has been shown how the pinch technique
gauge boson self--energies may be summed in Dyson series, in
exactly the same way as the conventional gauge--dependent
self--energies. We can therefore follow the standard diagonalization
procedure \cite{diag} except for the use
of PT self--energies in place
of the conventional functions.
 
The PT bare neutral current two--point functions can be 
written in matrix form as
\be\label{GammaNC}
\hat{\Gamma}_{NC} =
\left( \begin{array}{cc}
q^2 + \hat{\Sigma}_{\gamma\gamma} & \hat{\Sigma}_{\gamma Z} \\
\hat{\Sigma}_{\gamma Z}         & q^{2} - M_{Z}^{2} + \hat{\Sigma}_{ZZ}
\end{array} \right)\,.
\ee
The PT two--point component of the neutral current
amplitude for the interaction between fermions
with charges $Q$, $Q'$ and isospins $I_{3}$, $I_{3}'$ is then given
in terms of the inverse of the matrix $\hat{\Gamma}_{NC}$ by the expression
\bea\label{NC2pt}
\lefteqn{
\left( \begin{array}{cc}
eQ' & {\displaystyle \frac{e}{\sw\cw}}(I_{3}' - \sw^{2}Q')
\end{array} \right)
\hat{\Gamma}_{NC}^{-1}
\left( \begin{array}{c}
eQ \\ {\displaystyle \frac{e}{\sw\cw}}(I_{3} - \sw^{2}Q)
\end{array} \right) \,\,\,= }\nn \\
\label{born}
&\!\!\!\left( \begin{array}{cc}
eQ'  & {\displaystyle \frac{e}{\sw\cw}}(I_{3}' - \swe^{2}(q^{2})Q')
\end{array} \right)
\left( \begin{array}{cc}
\hat{\Delta}_{\gamma}(q^2)             & 0 \\ 
\phantom{\displaystyle \frac{e}{\sw}}0
\phantom{\displaystyle \frac{e}{\sw}}  & \hat{\Delta}_{Z}(q^2)
\end{array} \right)
\left( \begin{array}{c}
eQ \\ {\displaystyle \frac{e}{\sw\cw}}(I_{3} - \swe^{2}(q^{2})Q)
\end{array} \right)\,.\,\,\,\,\,\,
\eea
The r.h.s.\ of this equation, where the neutral current interaction
between the fermions has been written in diagonal
(i.e.\ Born--like) form, defines the diagonal propagator functions
$\hat{\Delta}_{\gamma}$ and $\hat{\Delta}_{Z}$ and
the effective weak mixing angle $\swe^{2}$.
They are given by
\bea\label{PTDeltagamma}
\hat{\Delta}_{\gamma}(q^2) &=& \frac{1}{q^{2} +
\hat{\Sigma}_{\gamma\gamma}(q^2)}\,,
\\
\label{PTDeltaZ}
\hat{\Delta}_{Z}(q^2)      &=&
\frac{1}{q^{2} -M_{Z}^{2} + \hat{\Sigma}_{ZZ}(q^2)
- \hat{\Sigma}_{\gamma Z}^{2}(q^2)/(q^{2} +
\hat{\Sigma}_{\gamma\gamma}(q^2))}\,,
\eea
and
\be\label{sweff}
\swe^{2}(q^2) = \sw^{2}\Biggl(1 +
\frac{\cw}{\sw}\frac{\hat{\Sigma}_{\gamma Z}(q^2)}
{q^{2}+\hat{\Sigma}_{\gamma\gamma}(q^2)} \Biggr)
= \swR^{2}\Biggl(1 +
\frac{\cwR}{\swR}\frac{\hat{\Sigma}_{R,\gamma Z}(q^2)}
{q^{2}+\hat{\Sigma}_{R,\gamma\gamma}(q^2)} \Biggr)\,.
\ee
In App.~B, it is shown how the amplitude (\ref{NC2pt}) 
for the Dyson--summed PT bare two--point component
of the neutral current interaction between fermions
takes exactly the same
form when expressed in terms of renormalized quantities.
This is a direct result of the
tree--level--like Ward identities obeyed by the PT one--loop
$n$--point functions.
The effective weak mixing angle $\swe^{2}$,
defined above in terms of bare quantities
in the first equality in (\ref{sweff}),
therefore takes exactly the same form when expressed
in terms of the corresponding renormalized quantities 
as in the second equality in (\ref{sweff}).
It is emphasized that the expressions
Eqs.~(\ref{PTDeltagamma})--(\ref{sweff}) are valid to all orders
in perturbation theory (although, to date, the PT proper self--energies
have only been computed to one--loop order).
 
The amplitude for the Dyson--summed PT two--point
component of the charged current interaction 
between fermions may be written
\be\label{CC2pt}
\frac{e^{2}}{2\sw^{2}}\Bigl(I_{+}I_{-}' + I_{-}I_{+}'\Bigr)
\hat{\Delta}_{W}(q^2)
\ee
where $I_{+}, I_{-}$ are the SU(2)${}_{L}$ isospin charge raising and 
lowering operators, and the propagator function is given by
\be\label{PTDeltaW}
\hat{\Delta}_{W}(q^2) = \frac{1}{q^{2} -M_{W}^{2} + \hat{\Sigma}_{WW}}\,.
\ee
In App.~B, the amplitude (\ref{CC2pt}) is also shown to take exactly 
the same form when expressed in terms of renormalized quantities.

We see that at the one--loop level, the only effect of the
$\hat{\Sigma}_{\gamma Z}$ self--energy is to correct the weak mixing angle.
At the one--loop level, the complex pole positions
$\bar{s}_{\gamma}$, $\bar{s}_{Z}$ and $\bar{s}_{W}$
of the diagonal propagators
$\hat{\Delta}_{\gamma}$, $\hat{\Delta}_{Z}$  and $\hat{\Delta}_{W}$
are thus given by the solutions of
\be
{\phantom{i = \gamma,Z,W.}}\hspace{25pt}
\bar{s}_{i} - M_{i}^{2} + \hat{\Sigma}_{ii}(\bar{s}_{i}) = 0\,,
\hspace{25pt}i = \gamma,Z,W,
\ee
(no sum over $i$; for the photon, $M_{\gamma} = 0$ 
and $\hat{\Sigma}_{\gamma\gamma}(0) = 0$ so that $\bar{s}_{\gamma} = 0$.)
The bare one--loop self--energy functions
$\hat{\Sigma}_{\gamma\gamma}$, $\hat{\Sigma}_{ZZ}$
and $\hat{\Sigma}_{WW}$
may be expanded around the corresponding pole positions
$\bar{s}_{\gamma}$, $\bar{s}_{Z}$ and $\bar{s}_{W}$ as\footnote{
For a discussion of expansions of the Dyson--summed electroweak 
self--energies about the pole positions, see Ref.~\cite{stuart}.}
\be
{\phantom{i = \gamma,Z,W.}}\hspace{25pt}
\hat{\Sigma}_{ii}(q^2) =
\hat{\Sigma}_{ii}(\bar{s}_{i}) + (q^{2}-\bar{s}_{i})\hat{\Pi}_{ii}(q^2)\,,
\hspace{25pt}i = \gamma,Z,W.
\ee
The diagonal propagators
$\hat{\Delta}_{\gamma}$, $\hat{\Delta}_{Z}$  and $\hat{\Delta}_{W}$
can then be written in terms of the functions
$\hat{\Pi}_{\gamma\gamma}$, $\hat{\Pi}_{ZZ}$ and $\hat{\Pi}_{WW}$ as
\be
{\phantom{i = \gamma,Z,W.}}\hspace{25pt}
\hat{\Delta}_{i}(q^2) = \frac{1}{(q^{2}-\bar{s}_{i})(1 +
\hat{\Pi}_{ii}(q^2))}\,,
\hspace{25pt}i = \gamma,Z,W,
\ee
i.e.\ with the pole explicitly factored out in each case.
The Dyson series of one--loop electroweak
radiative corrections included in these
propagators may then be fully accounted for by the three
effective charges
\bea\label{PTalphaeff}
\alpha_{\rm eff}(q^2)
&=&
\frac{e^{2}}{4\pi}
\,\frac{1}{1 + \hat{\Pi}_{\gamma\gamma}(q^2)}\,\,\,\,\,\,
\,\,\,\,\,\,\,\,\,\,\,=\,\,\,
\frac{e_{R}^{2}}{4\pi}\,\frac{1}{1 + \hat{\Pi}_{R,\gamma\gamma}(q^2)}\,, \\
\label{PTalphaZeff}
\alpha_{Z,{\rm eff}}(q^2)
&=&
\frac{e^{2}}{4\pi\sw^{2}\cw^{2}}\,\frac{1}{1+\hat{\Pi}_{ZZ}(q^2)}
\,\,\,=\,\,\,
\frac{e_{R}^{2}}{4\pi\swR^{2}\cwR^{2}}\,\frac{1}{1+\hat{\Pi}_{R,ZZ}(q^2)}\,,\\
\label{PTalphaWeff}
\alpha_{W,{\rm eff}}(q^2)
&=&
\,\,
\frac{e^{2}}{4\pi\sw^{2}}\,\frac{1}{1 + \hat{\Pi}_{WW}(q^2)}\,\,
\,\,\,=\,\,\,
\frac{e_{R}^{2}}{4\pi\swR^{2}}\,\frac{1}{1 + \hat{\Pi}_{R,WW}(q^2)}\,.
\eea
In Eqs.~(\ref{PTalphaeff})--(\ref{PTalphaWeff}),
the effective charges have also been written
in terms of the corresponding renormalized quantities.
Again, this is a direct result of the tree--level--like Ward identities
obeyed by the PT one-loop $n$--point functions, as described in App.~B.
In the on--shell scheme, the subtraction points used in \eq{sigmaR}
to define the renormalized self--energy functions are
given by the corresponding pole positions.
In this scheme, the functions $\hat{\Pi}_{R,ii}$ are then related to the
functions $\hat{\Sigma}_{R,ii}(q^{2})$ by the simple expressions
\bea
\hat{\Sigma}_{R,ii}(q^{2}) 
&=&  (q^{2}-\bar{s}_{i})
\Bigl(\hat{\Pi}_{ii}(q^{2}) - \hat{\Pi}_{ii}(\bar{s}_{i})\Bigr) \\ 
\hspace{100pt}
&=&  (q^{2}-\bar{s}_{i})\,\hat{\Pi}_{R,ii}(q^{2})\,,
\hspace{50pt}i = \gamma,Z,W.
\eea

Thus, we see that the radiative corrections
to the tree level electroweak two--point functions
are fully accounted for by {\em four} independent effective
quantities: the effective weak mixing angle in Eq.~(\ref{sweff}), and
the three effective charges in
Eqs.~(\ref{PTalphaeff})--(\ref{PTalphaWeff}).
In an exactly similar way to the QED effective charge in \eq{QEDalphaeff},
these four electroweak effective quantities
are gauge--independent, since the self--energy functions
$\hat{\Sigma}_{\gamma Z}$,
$\hat{\Sigma}_{\gamma\gamma}$, $\hat{\Sigma}_{ZZ}$ and
$\hat{\Sigma}_{WW}$  are gauge--independent; 
and also manifestly renormalization scale--
and scheme--independent, since they may be expressed entirely in terms of
bare quantities. 
 
%%%%%%%%%%%%%%%%%%%%%%%%%%%%%%%%%%%
 
\setcounter{equation}{0}
 
\section{The Effective Charges and the Renormalization Group}
 
It is instructive to compare the effective charges constructed in the previous
section with the corresponding running quantities defined from the
renormalization group. In particular, we will show how at high energy
the four independent effective quantities
$\alpha_{\rm eff}$, $\alpha_{Z,{\rm eff}}$,
$\alpha_{W,{\rm eff}}$ and $\swe^{2}$
in Eqs.~(\ref{PTalphaeff})--(\ref{PTalphaWeff}) and (\ref{sweff}) may be 
expressed
in terms of just two independent running quantities $\alphab$ and $\swb^{2}$.
 
The two $\beta$--functions associated with the electroweak
gauge groups SU(2)${}_{L}$ and U(1)${}_{Y}$ are defined in terms of the
corresponding renormalized couplings $\gR$ and $\gpR$, respectively, by
\bea\label{betadef}
\beta(\gR,\gpR) &=& \frac{\mu}{\gR^{2}}\frac{d\gR^{2}}{d\mu}
\,\,\,=\,\,\, \beta_{1}\frac{\gR^{2}}{4\pi^{2}} \,+\, \ldots \, ,\\
\label{betapdef}
\beta'(\gR,\gpR) &=& \frac{\mu}{\gpR^{2}}\frac{d\gpR^{2}}{d\mu}
\,\,\,=\,\,\, \beta_{1}'\frac{\gpR^{2}}{4\pi^{2}} \,+\, \ldots \, ,
\eea
where $\mu$ is the renormalization scale.
The first coefficients in the perturbative expansions are given by
\bea\label{beta1}
\beta_{1}  &=&
-\frac{43}{12}\,+\,\frac{2}{3}N_{g} \, , \\
\label{betap1}
\beta_{1}' &=& +\frac{1}{12} \,+\, \frac{10}{9}N_{g}\, ,
\eea
for $N_{g}$ fermion generations.
The term $+\frac{1}{12}$ in the expression for $\beta_{1}'$ is the
contribution of the complex Higgs doublet of scalar fields, while the
term $-\frac{43}{12} = -\frac{11}{6}\times 2 + \frac{1}{12}$
in the expression for $\beta_{1}$ is the sum of the contributions of the
pure SU(2)${}_{L}$ gauge fields and the Higgs doublet.
The solutions of the two differential equations in
Eqs.~(\ref{betadef}) and (\ref{betapdef}) then
define the SU(2)${}_{L}$ and 
U(1)${}_{Y}$ {\em running couplings} $\gb^{2}$ and $\gpb^{2}$, respectively:
in the leading--logarithm approximation,
\bea
\gb^{2}(\mu^{2})  &=& \frac{\gR^{2}}
{1 - \frac{\gR^{2}}{4\pi^{2}}\beta_{1}\frac{1}{2}\ln(\mu^{2}/\mu_{0}^{2})}\,,
\\
\gpb^{2}(\mu^{2}) &=& \frac{\gpR^{2}}
{ 1 - \frac{\gpR^{2}}{4\pi^{2}}\beta_{1}'\frac{1}{2}\ln(\mu^{2}/\mu_{0}^{2})}
\, ,
\eea
where $\gb^{2}(\mu_{0}^{2}) = \gR^{2}$ and $\gpb^{2}(\mu_{0}^{2}) =
\gpR^{2}$.
 
The two renormalized couplings $\gR$ and $\gpR$ may be traded for
a renormalized electromagnetic coupling $\alphaR$ and a renormalized
weak mixing angle $\swR$, defined by
\bea\label{alphaRdef}
\alphaR  &=& \frac{1}{4\pi}\frac{\gR^{2}\gpR^{2}}{\gR^{2}+\gpR^{2}}\,, \\
\label{swRdef}
\swR^{2} &=& \frac{\gpR^{2}}{\gR^{2}+\gpR^{2}}\,.
\eea
These definitions are by analogy with those of the tree level
electromagnetic coupling $\alpha$ and tree level weak mixing angle $\sw$
in terms of the bare couplings $g$ and $g'$.
The corresponding renormalization group functions $\beta^{\rm em}$ and
$\delta^{\sw}$ associated with $\alphaR$ and $\swR$ are then defined by
\bea\label{betaemdef}
\beta^{\rm em}(\alphaR,\swR) &=&
\,\,\frac{\mu}{\alphaR}\frac{d\alphaR}{d\mu}\,\,
\,\,\,=\,\,\,\beta_{1}^{\rm em}\frac{\alpha_{R}}{\pi} + \ldots \,, \\
\label{deltadef}
\delta^{\sw}(\alphaR,\swR) &=&
-\mu\,\frac{d\swR^{2}}{d\mu}
\,\,\,=\,\,\,\delta_{1}^{\sw}\frac{\alpha_{R}}{\pi} \,+\, \ldots \, ,
\eea
(the minus sign in \eq{deltadef} is for convenience).
The function $\beta^{\rm em}$ is thus the $\beta$--function for the
electromagnetic coupling in the presence of the electroweak
interactions. From the definitions Eqs.~(\ref{betadef}), (\ref{betapdef})
and (\ref{alphaRdef}), (\ref{swRdef}),
the functions $\beta^{\rm em}$ and $\delta^{\sw}$ are
given in terms of $\beta$ and $\beta'$ by
\bea
\beta^{\rm em} &=& \beta\swR^{2} \,+\, \beta'\cwR^{2}\,, \\
\delta^{\sw}   &=& \Bigl(\beta - \beta'\Bigr)\swR^{2}\cwR^{2}\,,
\eea
where $\cwR^{2} = 1 - \swR^{2}$.
The first coefficients in the perturbative expansions
of $\beta^{\rm em}$ and $\delta^{\sw}$ are thus given by
\bea
\beta_{1}^{\rm em} &=&
\beta_{1} + \beta_{1}'
\phantom{\beta_{1} - \swR^{2}\Bigl(\Bigr)}\,\,\,=\,\,\,
-\frac{7}{2}\,+\, \frac{16}{9}N_{g}\,, \\
\delta_{1}^{\sw}   &=&
\beta_{1} - \swR^{2}\Bigl(\beta_{1} + \beta_{1}'\Bigr) \,\,\,=\,\,\,
-\frac{43}{12} \,+\, \frac{2}{3}N_{g}
\,-\,\swR^{2}\biggl(-\frac{7}{2}\,+\, \frac{16}{9}N_{g} \biggr)\,.
\eea
The bosonic contributions to $\beta_{1}^{\rm em}$ and $\delta_{1}^{\sw}$
are precisely the coefficients
which appear in the $e^{+}e^{-}\ra W^{+}W^{-}$ tree level
cross section components
$\hat{\sigma}_{\gamma\gamma}$ and $\hat{\sigma}_{\gamma Z}$, respectively,
in the high energy limit $s/M_{W}^{2}\ra \infty$
[cf.\ Eqs.~(\ref{sigmagammagamma}) and (\ref{sigmagammaZ}), and the
third remark which follows them] and hence which appear
in the corresponding $W^{+}W^{-}$ contributions to the one loop
self--energies $\hat{\Sigma}_{\gamma\gamma}$ and $\hat{\Sigma}_{\gamma Z}$
[cf.\ Eqs.~(\ref{SigmaRggWW}) and (\ref{SigmaRgZWW})].
The {\em running} electromagnetic coupling $\alphab$ and the
{\em running} weak mixing angle $\swb^{2}$ are then defined as
the solutions of the differential equations in Eqs.~(\ref{betaemdef}) and
(\ref{deltadef}) or, equivalently, directly from
the running couplings $\gb^{2}$ and $\gpb^{2}$:
\bea
\alphab(\mu^{2})  &=&
\frac{1}{4\pi}\frac{\gb^{2}(\mu^{2})\gpb^{2}(\mu^{2})}
{\gb^{2}(\mu^{2})+\gpb^{2}(\mu^{2})}
\,\,\,=\,\,\,\frac{\alphaR}{1 -
\frac{\alphaR}{\pi}\beta_{1}^{\rm em}\frac{1}{2}\ln(\mu^{2}/\mu_{0}^{2})}\,,
\\
\swb^{2}(\mu^{2}) &=&
\frac{\gpb^{2}(\mu^{2})}{\gb^{2}(\mu^{2})+\gpb^{2}(\mu^{2})}
\phantom{\frac{1}{4\pi}}
\,\,\,=\,\,\,\swR^{2} \,-\,
\frac{\frac{\alphaR}{\pi}\delta_{1}^{\sw}\frac{1}{2}\ln(\mu^{2}/\mu_{0}^{2})}
{1 - \frac{\alphaR}{\pi}\beta_{1}
^{\rm em}\frac{1}{2}\ln(\mu^{2}/\mu_{0}^{2})}\,,
\eea
where the explicit expressions for $\alphab$ and $\swb^{2}$ are
again in the leading--logarithm approximation, with
$\alphab(\mu_{0}^{2}) = \alphaR$ and
$\swb^{2}(\mu_{0}^{2}) = \swR^{2}$\,.
 
{}From the definitions Eqs.~(\ref{PTalphaeff})--(\ref{PTalphaWeff}),
and \eq{sweff}, together with the explicit expressions
Eqs.~(\ref{SigmaRggWW})--(\ref{SigmaRZZWW}) and 
Eqs.~(\ref{SigmaRZZZH})--(\ref{SigmaRWWWgamma}) for the electroweak gauge field
contributions to the self--energies, and also the standard
fermion and Higgs contributions (not considered here),
we obtain:
\bea
\lim_{-q^{2}/M_{W}^{2}\ra\infty} \alpha_{\rm eff}(q^2)     & = &
\alphab(q^{2})\,,\phantom{\frac{q^{2}}{q^{2}}} \\
\lim_{-q^{2}/M_{W}^{2}\ra\infty} \alpha_{Z,{\rm eff}}(q^2) & = &
\frac{\alphab(q^{2})}{\swb^{2}(q^{2})\cwb^{2}(q^{2})}\,, \\
\lim_{-q^{2}/M_{W}^{2}\ra\infty} \alpha_{W,{\rm eff}}(q^2) & = &
\frac{\alphab(q^{2})}{\swb^{2}(q^{2})}\,, \\
\lim_{-q^{2}/M_{W}^{2}\ra\infty} \swe^{2}(q^{2})               & = &
\swb^{2}(q^{2})\,.\phantom{\frac{q^{2}}{q^{2}}}
\eea
Thus, in the high energy limit, when all masses can
be neglected, the $q^2$--dependence of the four effective quantities
$\alpha_{\rm eff}$, $\alpha_{Z,{\rm eff}}$,
$\alpha_{W,{\rm eff}}$ and $\swe^{2}$
is fully specified by the renormalization group running of the two
independent quantities $\alphab$ and $\swb^{2}$.
 
%%%%%%%%%%%%%%%%%%%%%%%%%%%%%%%%%%%%%%%%%%%%%%%%%%
 
\pagebreak

\setcounter{equation}{0}
\setcounter{footnote}{0}
 
\section{The Effective Charges and the Background Field Method}

In this section we compare the effective charges constructed in Sec.~5
with those obtained directly from one--loop perturbation theory in the
background field method (BFM). This comparison is pertinent for two reasons.
First, when a non--abelian theory is quantized in the BFM,
the background gauge boson self--energies capture fully
the renormalization group running of the gauge couplings, just as in QED.
The BFM then naturally provides a well--defined field theoretic framework 
in which the renormalization group
running couplings may be extended into the non--asymptotic 
region to define sets of background field effective charges.
Second, as is now well known, there exists a connection between the
one--loop PT $n$--point functions and those obtained in the BFM.
In Secs.~3 and 4, it was shown that when the QED relation between
one--loop contributions to the gauge field self--energy and the 
tree level cross sections for the corresponding physical processes 
is extended to the electroweak theory, 
the gauge field contributions to the electroweak self--energies which result 
are identical to those obtained in the PT. 
It is therefore instructive to reconsider the connection between the PT and 
the BFM in the context of the approach followed here.

In the BFM \cite{abbott}, 
the bosonic fields are decomposed into background and quantum
components, and the gauge fixing for the quantum gauge fields
then chosen such that the effective action remains explicitly invariant
under gauge transformations of the background fields.
As a result of this exact background gauge invariance,
the one-particle-irreducible background field $n$--point functions,
although still dependent on the quantum gauge fixing parameter $\xi_{Q}$,
obey simple QED--like Ward identities to all orders in perturbation theory.
In particular, in the BFM formulation of the electroweak theory 
\cite{denner1}, the renormalization constants for the background
$\gamma$, $Z$ and $W$ gauge fields are $\xi_{Q}$--independent, 
and are related to those of the electromagnetic coupling and the 
weak mixing angle precisely as in Eqs.~(\ref{Zids}).
Consequently, the set of effective charges defined by the Dyson summation
of the background gauge field self--energies are renormalization scale-- 
and scheme--independent, and at asymptotic energies match on to the
corresponding running quantities defined from the renormalization group.
However, in the non--asymptotic region the BFM effective charges
retain a dependence on the quantum gauge parameter $\xi_{Q}$.
Thus, the effective charges defined via the Dyson summation of the BFM
self--energies, although well--defined at the field theoretic level
and while satisfying the constraints of the renormalization group, 
are not unique.

It has been observed by various authors~\cite{denner1,hashimoto,edrnjw}
that the PT gauge--independent
one--loop $n$--point functions {\em coincide} with the background field
$n$-point functions computed in the Feynman quantum gauge $\xi_{Q} = 1$,
both in QCD and the electroweak Standard Model.\footnote{
For an extensive discussion of the connection between the PT and the BFM
in perturbation theory at the one--loop level and beyond,
see Ref.~\cite{pilaftsis}.}
Furthermore, it has been argued \cite{denner1,denner2}
that the PT $n$--point functions
are not distinguished on physical grounds from the BFM $n$--point functions
computed for an arbitrary value of $\xi_{Q}$.
In the case of the gauge boson self--energies, this argument
leads to the conclusion that, away from the asymptotic region
governed by the renormalization group, 
there is no unique, physically distinguished
way in which to extend the concept of an effective charge from QED 
to the electroweak sector of the Standard Model in particular, 
and to non--abelian gauge theories in general.

However, as emphasized in Ref.~\cite{papapil2}, for $\xi_{Q}\neq 1$
the imaginary parts of the BFM electroweak self--energies
$\Sigma_{ij}^{\scriptscriptstyle {\rm BFM}}$, $i,j = \gamma,Z,W$,
include terms with unphysical thresholds.
For example, for the one--loop contributions of the $W$ and its associated
would--be Goldstone boson and ghost to 
$\Sigma_{ZZ}^{\scriptscriptstyle {\rm BFM}}$~\cite{denner2}, one obtains
\bea
\lefteqn{
\im \Sigma_{ZZ}^{\scriptscriptstyle {\rm BFM}(WW)}(\xi_{Q},q^{2})
\,\,=\,\, \im \hat{\Sigma}_{ZZ}^{\scriptscriptstyle (WW)}(q^{2}) 
\,\,+\,\, \frac{\alpha}{24\sw^{2}\cw^{2}} (q^{2} - M_{Z}^{2})
\frac{1}{M_{Z}^{4}} \,\times  } \nn \\
& &
\biggl\{ \Bigl((8M_{W}^{2} + q^{2})(M_{Z}^{2} + q^{2})
+ 4M_{W}^{2}(4M_{W}^{2} + 3M_{Z}^{2} + 2q^{2})  \Bigr) 
\lambda^{\frac{1}{2}}(q^{2},M_{W},M_{W})
\vartheta(q^{2} - 4M_{W}^{2}) \nn \\
& &
+\Bigl((8M_{W}^{2} + q^{2})(M_{Z}^{2} + q^{2})
- 4M_{W}^{2}(4M_{W}^{2} + 3M_{Z}^{2} + 2q^{2}) \nn \\
& &
-4(\xi_{Q}-1)M_{W}^{2}(4M_{W}^{2} + M_{Z}^{2} + q^{2}) \Bigr) 
\lambda^{\frac{1}{2}}(q^{2},\sqrt\xi_{Q}M_{W},\sqrt\xi_{Q}M_{W})
\vartheta(q^{2} - 4\xi_{Q}M_{W}^{2}) \nn \\
& &
-2\Bigl(8M_{W}^{2} + q^{2} -2(\xi_{Q}-1)M_{W}^{2}
+(\xi_{Q}-1)^{2}M_{W}^{4}q^{-2}\Bigr)(M_{Z}^{2} + q^{2}) \times \nn \\
& &
\lambda^{\frac{1}{2}}(q^{2},M_{W},\sqrt\xi_{Q}M_{W})
\vartheta(q^{2} - (1 + \sqrt\xi_{Q})^{2}M_{W}^{2}) \biggr\}\,,
\eea
where $\lambda(q^{2},M_{i},M_{j})$ was defined after \eq{sigmaZH}.
These gauge--dependent unphysical thresholds are artefacts of the
BFM electroweak $R_{\xi}$--like gauge fixing procedure,
and exactly cancel in the calculation of any physical process.
For example, one can consider in the framework of the BFM
the radiative corrections to the process $e^{+}e^{-} \ra e^{+}e^{-}$
in the case of forward scattering, as discussed in the introduction.
The contributions of the $W$ and its associated would--be Goldstone boson
and ghost to the one--loop BFM radiative corrections are just as
illustrated schematically on the l.h.s.\ of Fig.~3.
The unphysical thresholds which occur in the imaginary parts of the
various one--loop self--energy contributions are then exactly cancelled by 
corresponding unphysical contributions from the imaginary parts of the 
one--loop vertex and box contributions.
Thus, as stated in the introduction,
these cancellations leave just the contribution proportional to
the tree level cross section for the on--shell physical process 
$e^{+}e^{-} \ra W^{+}W^{-}$, illustrated on the r.h.s.\ of Fig.~3,
with threshold at $q^{2} = 4M_{W}^{2}$.

Clearly, the unphysical thresholds which occur in the
imaginary parts of the electroweak $n$--point functions in the BFM
(or indeed with any other gauge fixing procedure)
are not experimentally measurable quantities.
It follows that {\em the gauge field contributions
to the renormalized BFM electroweak self--energies
$\Sigma_{R,ij}^{\scriptscriptstyle {\rm BFM}}(\xi_{Q},q^{2})$
computed with $\xi_{Q} \neq 1$
cannot be reconstructed from the tree level cross sections
for the corresponding on--shell physical processes.}
Furthermore, as remarked in Ref.~\cite{denner2} (third paper),
it is not a priori obvious that, even for $\xi_{Q} = 1$, all thresholds
in the imaginary parts of the BFM self--energies are due to physical fields.
Here, by obtaining the full one--loop $W$ contribution to the
renormalized PT neutral current self--energies,
coincident with the BFM self--energies at $\xi_{Q} = 1$,
directly from the tree level cross section for the on--shell physical
process $e^{+}e^{-} \ra W^{+}W^{-}$,
we have shown explicitly that, in the BFM at $\xi_{Q} = 1$,
the thresholds which occur at $q^{2} = 4M_{W}^{2}$ 
are due solely to the physical $W^{+}W^{-}$ pair.
We therefore conclude that the particular value $\xi_{Q} = 1$ in the BFM
precisely {\em is} distinguished on physical grounds from
all other values of $\xi_{Q}$.\footnote{
For a discussion of the pathologies which result from the Dyson
summation of self--energies which include unphysical thresholds,
see Ref~\cite{papapil2}.}

%%%%%%%%%%%%%%%%%%%%%%%%%%%%%%%%%%%%%%%%%%%%%%%%%%
 
\setcounter{equation}{0}
 
\section{Phenomenological Determination of the Effective Charges}

Finally, we turn to the extraction from experiment of the effective charges
and weak mixing angle.
In QED, the fine structure constant $\alpha$ together with the mass(es) of the
fermion(s) provides the experimental input required to determine the
parameters of the theory.
The overall scale of the effective charge is then determined by $\alpha$,
while in the on--shell scheme the subtraction point in the dispersion relation 
\eq{qeddisprel}
for the renormalized vacuum polarization function is at $s = 0$.
The one--loop 
contributions to the spectral function $\im\Pi$ are then directly 
proportional to components of the tree level cross sections for
the corresponding fermion scattering processes, as described in the
Introduction.

In the electroweak Standard Model, the parameters of the theory are
determined by three independent experimental inputs together with the
masses of the fermions and the Higgs boson. In the on--shell scheme,
these three parameters are chosen to be the fine structure constant,
defined from Compton scattering in the classical Thomson limit,
and the masses of the $W$ and $Z$ gauge bosons, defined from the pole
positions of the corresponding propagators.

At the one--loop level in the electroweak theory, the
quantum correction to the fine structure constant which appears in the
classical Compton scattering process is given by
\bea\label{ewcompton}
\alpha
&=&
\frac{e^{2}}{4\pi}\biggl( 1 - \Pi_{\gamma\gamma}(\xi,0)
-\frac{2\sw}{\cw}\frac{\Sigma_{\gamma Z}(\xi,0)}{M_{Z}^{2}}
\biggr) \\
\label{ewPTcompton}
&=&
\frac{e^{2}}{4\pi}\biggl( 1 - \hat{\Pi}_{\gamma\gamma}(0) \biggr)
\,\,\,=\,\,\,
\frac{e_{R}^{2}}{4\pi}\biggl( 1 - \hat{\Pi}_{R,\gamma\gamma}(0) \biggr)\,.
\eea
In \eq{ewcompton}, $\Pi_{\gamma\gamma}(\xi,q^2)$ and
$\Sigma_{\gamma Z}(\xi,q^2)$ are the conventional
gauge--dependent photon vacuum polarization
and $\gamma$--$Z$ self--energy in the class of renormalizable
($R_{\xi}$) gauges with (common) gauge parameter $\xi$.
At vanishing four--momentum transfer $q^2 = 0$, the combination
of these two functions in \eq{ewcompton}
is gauge--independent and is precisely equal to the the PT function
$\hat{\Pi}_{\gamma\gamma}(q^{2})$ at $q^{2}= 0$.~\footnote{
The expressions for the PT functions
$\hat{\Pi}_{\gamma\gamma}(q^2)$ and $\hat{\Sigma}_{\gamma Z}(q^2)$ in terms of
$\Pi_{\gamma\gamma}(\xi=1,q^2)$ and $\Sigma_{\gamma Z}(\xi=1,q^2)$ can be
found in Eqs.~(16a) and (16b) of Ref.~\cite{degsir}. Using these
expressions, together with the fact that $\hat{\Sigma}_{\gamma Z}(0) =0$,
one obtains the first equality in (\ref{ewPTcompton}).}
Thus, {\em the PT photon vacuum polarization
specifies fully the one--loop electroweak radiative corrections
to the classical Compton scattering process in the Thomson limit}.
This is precisely analogous to QED.
In the second equality in (\ref{ewPTcompton}), the Ward identity
\eq{Deltagammarenorm} has been used to write the one--loop expression
for $\alpha$ in terms of renormalized quantities.
In the on--shell scheme, with subtraction point $s_{\gamma\gamma} = 0$,
$\hat{\Pi}_{R,\gamma\gamma}(0) = 0$ and
$e_{R}$ is identified with the physical electron charge.

Also at the one--loop level, the pole positions
$\bar{s}_{Z}$ and $\bar{s}_{W}$ of the conventional
$R_{\xi}$ gauge $Z$ and $W$ propagators are given by
\bea\label{convpole}
\hspace{20pt}
\bar{s}_{i} &=& M_{i}^{2} - \Sigma_{ii}(\xi,M_{i}^{2}) \\
\label{PTpole}
            &=& M_{i}^{2} - \hat{\Sigma}_{ii}(M_{i}^{2})
  \,\,\,=\,\,\, M_{R,i}^{2} + \delta M_{i}^{2}  
- \hat{\Sigma}_{ii}(M_{i}^{2})\,, \hspace{20pt} i= Z,W.
\eea
At $q^{2} = M_{Z}^{2}$ [$M_{W}^{2}]$, the conventional
$R_{\xi}$ gauge self--energy
function $\Sigma_{ZZ}(\xi,q^{2})$ [$\Sigma_{WW}(\xi,q^{2})$]
is gauge--independent and
is precisely equal to the PT function
$\hat{\Sigma}_{ZZ}(q^{2})$ [$\hat{\Sigma}_{WW}(q^{2})$] 
at $q^{2} = M_{Z}^{2}$ [$M_{W}^{2}]$.
Thus, {\em the PT self--energies do not shift the position of the
complex poles}. This was shown to one-loop order in 
Refs.~\cite{PTmassive,degsir} and to higher orders in Ref.~\cite{papapil1}.
In the second equality in (\ref{PTpole}), the bare masses have been
written in terms of renormalized masses and counterterms. In the
on--shell scheme, with renormalization conditions
$\delta M_{i}^{2} = \re \hat{\Sigma}_{ii}(M_{i}^{2})$, $i = Z,W$,
$M_{R,Z}$ and $M_{R,W}$ are identified with
the physical $Z$ and $W$ masses. 
In this scheme, the renormalized weak mixing angle is then defined by
\be
\cwR^{2} = 1 - \swR^{2} = \frac{M_{R,W}^{2}}{M_{R,Z}^{2}}\,.
\ee

Thus, in the on--shell scheme at the one--loop level,
the overall scale of each of the
three effective charges 
Eqs.~(\ref{PTalphaeff})--(\ref{PTalphaWeff}) 
and the effective weak mixing angle Eq.~(\ref{sweff}) 
is given by the fine structure constant $\alpha$
and the ratio $\cwR$ of the physical $W$ and $Z$ masses; 
the subtraction points $s_{\gamma\gamma}$ and $s_{\gamma Z}$
for $\hat{\Sigma}_{R,\gamma\gamma}$ and $\hat{\Sigma}_{R,\gamma Z}$
in the dispersion relation \eq{sigmadisprel} are both at zero; 
and the subtraction points $s_{ZZ}$ and $s_{WW}$ 
for $\hat{\Sigma}_{R,ZZ}$ and $\hat{\Sigma}_{R,WW}$ in the
dispersion relation \eq{sigmadisprel} are given by the squares of the
physical $Z$ and $W$ masses, respectively.

It remains to extract from experiment the absorptive parts 
$\im\hat{\Sigma}_{\gamma\gamma}$, $\im\hat{\Sigma}_{\gamma Z}$, 
$\im\hat{\Sigma}_{ZZ}$ and
$\im\hat{\Sigma}_{WW}$ of the electroweak self--energy functions.
In QED, the one--loop muon contribution of to the spectral function
$\im\Pi$ is determined directly from the tree level cross section
$\sigma(e^{+}e^{-}\rightarrow\mu^{+}\mu^{-})$. 
For the one--loop electron contribution to $\im\Pi$, however,
it is necesssary to project out the self--energy--like
component of the tree level cross section
$\sigma(e^{+}e^{-}\rightarrow e^{+}e^{-})$. 
This is possible due to the linear independence of the
self--energy--, vertex-- and box--like components
of the Bhabha differential cross section \eq{dsigmaff}.
Similarly, in the electroweak Standard Model, in order to obtain
the gauge boson contributions to the functions
$\im\hat{\Sigma}_{\gamma\gamma}$, $\im\hat{\Sigma}_{\gamma Z}$, 
$\im\hat{\Sigma}_{ZZ}$ and
$\im\hat{\Sigma}_{WW}$, it is necessary to project out
the self--energy--like components of the physical cross sections  
$e^{+}e^{-} \rightarrow W^{+}W^{-}$ and
$e^{+}e^{-} \rightarrow ZH$ 
for the neutral current functions, and
$e^{+}\nu_{e} \rightarrow W^{+}Z$,
$e^{+}\nu_{e} \rightarrow W^{+}\gamma$ and 
$e^{+}\nu_{e} \rightarrow W^{+}H$
for the charged current function.
These projections can be obtained by the appropriate convolution of the full
differential cross sections with specific angular functions. To illustrate
the procedure, we shall explicitly discuss the simplified limit
in which the weak mixing angle is set to zero,
i.e.\ the case of a broken SU(2)${}_{L}$ gauge theory.
In this case, it is sufficient to know the differential cross sections in
order to make the projection. The general case with $\sw^{2}\neq 0$ 
requires in addition the observation of spin density matrices~\cite{renard};
though technically more involved, the procedure is in principle the same.
The important property which we wish to stress is that, 
in contrast to the conventional gauge--dependent self--energies,  
the absorptive parts $\im\hat{\Sigma}_{\gamma\gamma}$,
$\im\hat{\Sigma}_{\gamma Z}$, 
$\im\hat{\Sigma}_{ZZ}$  and $\im\hat{\Sigma}_{WW}$ 
of the pinch technique self--energies
are directly related to components of physical 
cross sections which are, in principle, experimentally observable.

We thus set $\sw^{2}=0$ in Eqs.~(\ref{Mgammagamma})--(\ref{Mnunu}),
with $\alpha = g^{2}\sw^{2}/4\pi$. 
It is also convenient to introduce the variables
\be
x = \cos\theta\,, \hspace{50pt}
z = \frac{1 + \beta^{2}}{2\beta}\,.
\ee
Then
\be
\frac{s}{t} =- \frac{2}{\beta}\frac{1}{(z - x)}\,.
\ee
The function $\hat{M}_{\nu\nu}$ \eq{Mnunu} specifying the
box--like contribution to the differential cross section
has a double pole at the (unphysical) point $z = x$;
the function $\hat{M}_{Z\nu}$ \eq{MZnu} specifying the
vertex--like contribution has a single pole at $z = x$;
and the function $\hat{M}_{ZZ}$ \eq{MZZ} specifying the
self--energy--like contribution has no pole.
If the differential cross section is multiplied by the factor
$(z-x)^{2}$, the resulting observable then has a simple
degree four polynomial dependence on the variable $x = \cos\theta$.
The product of the differential cross section and $(z-x)^{2}$
can therefore be expanded in terms of any set of five
linearly independent polynomials
$F_{i}(s,x)$, $i = 1,2\ldots 5$ of degree four in $x$, and
which may also depend on $s$\,:
\be\label{obs}
(z - x)^{2}
\left.\frac{d\sigma(e^{+}e^{-}\rightarrow W^{+}W^{-})}{dx}\right|_{\sw = 0}
=
\frac{g^{4}}{64\pi}\beta\frac{s}{(s-M^{2})^{2}}
\sum_{i=1}^{5}A_{i}(s)F_{i}(s,x)
\ee
(we have used $d\Omega = 2\pi\,dx$ and on the
r.h.s.\ extracted an overall factor for convenience).
 
We now make the following choice for the polynomials
$F_{i}(s,x)$, with from Eqs.~(\ref{dhatsigmaeeWW}),
(\ref{MZZ}), (\ref{MZnu}) and (\ref{Mnunu})
the corresponding coefficient functions $A_{i}(s)$:
\be\label{FiAi}
\begin{array}{rclrcl}
F_{1}(s,x) &=& (z-x)^{2} \hspace{60pt} &
A_{1}(s)   &=&{\displaystyle \frac{5}{32}(\beta^{2}-12)
\,\vartheta(s-4M^{2})} \,,\\
F_{2}(s,x) &=& (z-x)^{2}\,x^{2}        &
A_{2}(s)   &=&{\displaystyle -\frac{9}{32}\beta^{2}
\,\vartheta(s-4M^{2})} \,;\\ \\
F_{3}(s,x) &=& (z-x)(1-x^{2})          &
A_{3}(s)   &=& {\displaystyle -\frac{\beta}{2}\biggl(\frac{s-M^{2}}{s}\biggr)
\,\vartheta(s-4M^{2})} \,,\\
F_{4}(s,x) &=& (z-x)(1-\beta x)        &
A_{4}(s)   &=& {\displaystyle \frac{2}{\beta}\biggl(\frac{s-M^{2}}{s}\biggr)
\,\vartheta(s-4M^{2})} \,;\\ \\
F_{5}(s,x) &=& 1-x^{2}                 &
A_{5}(s)   &=& {\displaystyle \frac{1}{2}\biggr(\frac{s-M^{2}}{s}\biggr)^{2}
\vartheta(s-4M^{2})}\,.
\end{array}
\ee
The above choice of polynomials $F_{i}(s,x)$ is such as to isolate
explicitly the self--energy--like, vertex--like and
box--like components of the differential cross section. Thus,
the coefficients $A_{1}(s)$ and $A_{2}(s)$ contribute only to the
self--energy--like component $d\hat{\sigma}_{ZZ}/dx$;
the coefficients $A_{3}(s)$ and $A_{4}(s)$ contribute only to the
vertex--like component $d\hat{\sigma}_{Z\nu}/dx$;
and the coefficient $A_{5}(s)$ contributes only to the
box--like component $d\hat{\sigma}_{\nu\nu}/dx$.
In particular, the imaginary part of the
$W$ contribution to the self--energy is then given by
\be\label{imSigmaA1A2}
\left.\im\,\hat{\Sigma}_{R,ZZ}^{\scriptscriptstyle (WW)}(s)
\right|_{\sw = 0} =
\frac{g^{2}}{4\pi}\beta s\biggl( A_{1}(s) +
\frac{1}{3}A_{2}(s)\biggr)\,.
\label{SA}
\ee
The Wronskian for the five functions $F_{i}(s,x)$ is given by
$W(F_{i})= 288(1-z\beta)(1-z^{2})$. The functions are
therefore linearly independent for all values of $s$ except
at the zeros of $W(F_{i})$ occurring at
$z=\beta^{-1}$ and (equivalently) $z=1$, i.e.\ at $s/M_{W}^{2}\ra\infty$.
 
To project out the functions $A_{i}(s)$ we construct a further set of five
degree four polynomials
$\tilde{F}_{i}(s,x)$ satisfying the orthogonality conditions
\be
{\phantom{i,j = 1,2\ldots 5}}\hspace{25pt}
\int_{-1}^{1}dx\,F_{i}(s,x)\tilde{F}_{j}(s,x) = \delta_{ij}\,,
\hspace{25pt}i,j = 1,2\ldots 5\,.
\ee
The explicit expressions for the
$\tilde{F}_{i}(s,x)$ corresponding to the specific choice of
$F_{i}(s,x)$ in (\ref{FiAi}) are given in App.~C.
Using these functions $\tilde{F}_{i}(s,x)$, the coefficient functions
$A_{i}(s)$ may then be projected out from the observable formed from the
product of the differential cross section and the kinematic factor
$(z-x)^{2}$\,:
\be\label{proj}
\int_{-1}^{1}dx\,\tilde{F}_{i}(s,x)\,
(z-x)^{2}\left.\frac{d\sigma}{dx}\right|_{\sw = 0}
=
\frac{g^{4}}{64\pi}\beta\frac{s}{(s-M^{2})^{2}}\,A_{i}(s)\,.
\ee
Thus, by expanding the product of the differential cross section
and $(z-x)^{2}$ in terms of
functions $F_{i}(s,x)$ which characterize explicitly the
angular ($x$) dependence of the \mbox{self--energy--,} 
vertex-- and box--like components of the
differential cross section, it is possible to extract
$\im\,\hat{\Sigma}_{R,ZZ}^{\scriptscriptstyle (WW)}(s)|_{\sw = 0}$
directly from
$d\sigma(e^{+}e^{-}\rightarrow W^{+}W^{-})/dx |_{\sw = 0}$.
 
%%%%%%%%%%%%%%%%%%%%%%%%%%%%%%%%%%%%%%%%%%%%%%%%%%
 
\setcounter{equation}{0}
 
\section {Conclusions}

The analysis presented here demonstrates the existence of 
effective charges in non--abelian gauge
theories with properties precisely analogous to those of the well--known 
effective charge of QED. We have shown this by an explicit
construction in the neutral current sector of the electroweak Standard
Model, where the processes
$e^{+}e^{-}\ra W^{+}W^{-}$ and $e^{+}e^{-}\ra ZH$ play the
same r\^{o}le as the processes
$e^{+}e^{-}\ra f^{+}f^{-}$ in QED with fermions $f = e,\mu\ldots$~.

For the process $e^{+}e^{-}\ra W^{+}W^{-}$, we have shown how
the tree level cross section naturally decomposes into components
which are uniquely defined according to the propagator structure
occurring in the square of the modulus of the S--matrix element,
and which are {\em individually} well--behaved at high energy.
This decomposition follows directly from the systematic
use of tree level Ward identities
to implement cancellations among contributions
originating from diagrams with distinct $s$-- and $t$--dependence.
These cancellations
are purely non--abelian in character, and have no analogue in QED.
The resulting expressions for the self--energy--like components
of the tree level cross section explicitly display the $W$
contributions to the one--loop electroweak $\beta$--functions,
in the same way as the self--energy--like components of the
tree level $e^{+}e^{-}\ra f^{+}f^{-}$ cross sections display the
corresponding fermion contributions.
We have then used the self--energy--like components of the tree level
$e^{+}e^{-}\ra W^{+}W^{-}$ cross section and the full
tree level
$e^{+}e^{-}\ra ZH$ cross section
to obtain one--loop gauge boson contributions to renormalized
electroweak self--energy functions
using dispersion relations, by direct analogy with QED.
The self--energy functions so obtained are {\em identical} to the
corresponding functions obtained in the pinch technique.
These self--energy functions, together with that for the $W$,
were then used to construct
the electroweak effective charges and the effective weak mixing angle. 
We have shown how, in the on--shell renormalization scheme, the 
subtractions points in the dispersion relations
and also the overall scale of the effective charges and weak mixing angle
are related to the three experimental inputs used to
determine the basic parameters of the electroweak theory.
Furthermore, we have described how the gauge boson contributions
to the absorptive parts of the self--energy functions
may be projected out directly from the  
corresponding tree level physical cross section.

Radiative corrections to two--point functions in the electroweak Standard
Model are thus fully accounted for by {\em four}
independent functions: the three effective charges in
Eqs.~(\ref{PTalphaeff})--(\ref{PTalphaWeff}), and the effective weak mixing
angle in Eq.~(\ref{sweff}). These four effective quantities are
gauge--independent, and also renormalization scale-- and
scheme--independent.  In the high energy limit, when all masses can be
neglected, the $q^2$--dependence of these four effective quantities
is fully specified by the
renormalization group running of two independent quantities, which may be
chosen to be the running electromagnetic coupling
$\alphab$ and the running weak mixing angle $\swb^{2}$. 
At all other momentum scales where
masses cannot be neglected, the effective charges we have
constructed provide the unique and unambiguous extension of the QED
concept of an effective charge to include the 
contributions of massive gauge bosons. 
The comparison of these functions with experiment, along the lines
discussed in Ref.~\cite{hagiwara}, should provide a natural way
to parameterize possible deviations from the Standard Model 
due to ``new physics''.

We consider the fact that the
self--energy functions which we have constructed from the
physical processes $e^{+}e^{-}\ra W^{+}W^{-}$ and $e^{+}e^{-}\ra ZH$ turn
out to be identical to those obtained by the pinch technique provides
a convincing argument in favour of the pinch technique approach to the
construction of effective charges. Indeed, it is remarkable that
the simple QED unitarity relation among components of the tree level
cross section for the interaction of on--shell particles and the
imaginary parts of the corresponding one--loop self--energy,
vertex and box functions may be extended to non--abelian gauge theories
in this way. Furthermore, the cancellation mechanism illustrated
in Fig.~5 responsible for the good high energy behaviour
of the tree level $e^{+}e^{-} \ra W^{+}W^{-}$ cross section
is also that responsible for the gauge--independence of
the $W$ contributions to the PT self-energies
$\hat{\Sigma}_{\gamma\gamma}$, $\hat{\Sigma}_{\gamma Z}$
and $\hat{\Sigma}_{ZZ}$ obtained via the rearrangement
of one--loop perturbation theory diagrams.
Clearly, it would be worthwhile now to extend
this construction to the one--particle--irreducible two--loop level.
 
The electroweak example which we have worked out also gives support to the
QCD effective charge obtained by the pinch technique, recently
extensively discussed in ref.~\cite{njw2}, as the appropriate quantity to be
used in renormalon calculus in the one--loop approximation of the 
$\beta$--function. 
In a similar way to the electroweak effective charges constructed here,
the QCD effective charge constructed in \cite{njw2} corresponds
to a well--defined class of components
of Feynman diagrams selected by the tree level Ward identities.
 
In spite of the progress which has been made in understanding the 
concept of an
effective charge in non--abelian gauge theories, there remain issues which
we still would like to clarify:
 
\begin{itemize}
\item
One issue is the possible connection with
conformal invariance: once the higher orders of perturbation theory are
absorbed in the effective charge, one would expect, in the
massless limit, conformal invariance to be restored. What are then
the corresponding constraints?
\item
So far we only have a diagrammatic understanding of the construction
of an effective charge. Although, as explained in this paper, we are
able to relate this construction to physical observables, we still
lack a formal understanding, in particular in terms of a path integral
formulation. What is the constraint on the path integral which
projects the effective charge two--point functions directly?
\end{itemize}
 
\noi
We hope to come back to these questions in the near future.

\vspace{10pt}

It is a pleasure to thank J.~Bernab\'eu, F.~Boudjema, R.~Coquereaux, 
M.~Hassler, S.~Peris, M.~Perrottet and A.~Pich for useful discussions. 
NJW acknowledges the financial support of EC HCM grant ERB4001GT933989.

%%%%%%%%%%%%%%%%%%%%%%%%%%%%
%%%%%%%%%%%%%%%%%%%%%%%%%%%%

\pagebreak

\appendix
\setcounter{equation}{0}
\def\theequation{A.\arabic{equation}}

{\Large{\bf Appendix A}}

\vspace{10pt}

\noi
In this Appendix, we give the explicit expressions for the
renormalized self--energy functions
$\hat{\Sigma}_{R,ZZ}^{\scriptscriptstyle (ZH)}$,
$\hat{\Sigma}_{R,WW}^{\scriptscriptstyle (WH)}$,
$\hat{\Sigma}_{R,WW}^{\scriptscriptstyle (W\gamma)}$ and
$\hat{\Sigma}_{R,WW}^{\scriptscriptstyle (WZ)}$
defined in Sec.~4. To this end, it is convenient to introduce
the function $L_{ij}$ defined by
\be\label{Lij}
L_{ij}(q^{2}) =
\frac{1}{4}
\lambda_{ij}^{\frac{1}{2}}\ln \rho_{ij}
\,-1 \,
+\frac{1}{2}\ln \frac{2M_{i}M_{j}}{M_{i}^{2} + M_{j}^{2}}
+ \frac{1}{2}\kappa_{ij}\ln \frac{M_{i}}{M_{j}} \,,
\ee
where
\bea
\lambda_{ij} &=&
\biggl(1 - \frac{(M_{i}+M_{j})^{2}}{q^{2}}\biggr)
\biggl(1 - \frac{(M_{i}-M_{j})^{2}}{q^{2}}\biggr)\,, \\
\rho_{ij} &=&
\frac{M_{i}^{2} + M_{j}^{2} - (1+\lambda_{ij}^{\frac{1}{2}})q^{2}}
     {M_{i}^{2} + M_{j}^{2} - (1-\lambda_{ij}^{\frac{1}{2}})q^{2}} \,,\\
\kappa_{ij} &=& 
\frac{M_{i}^{2}-M_{j}^{2}}{q^{2}}\,.
\eea
Thus
\bea
\im L_{ij}(q^{2}) \!&=&\! -\frac{\pi}{2}\lambda_{ij}^{\frac{1}{2}}\,
\vartheta(q^{2} - (M_{i}\!+\!M_{j})^{2})\,, \\
q^{2}\frac{dL_{ij}(q^{2})}{dq^{2}} \!&=&\!
\frac{1}{\lambda_{ij}}\Biggl\{ 
\biggl( \frac{M_{i}^{2}\!+\!M_{j}^{2}}{q^{2}} -\kappa_{ij}^{2}\biggr)
\biggl(\frac{1}{4}\lambda_{ij}^{\frac{1}{2}}\ln \rho_{ij} \,-1 \biggr) 
+ \frac{1}{2}\biggl(1 - \kappa_{ij}^{2}\biggr) \Biggr\}
-\frac{1}{4}\kappa_{ij} \ln\frac{M_{i}}{M_{j}}\,.\hspace{25pt}
\eea
For the particular case $i=j=W$, $\lambda_{WW}^{\frac{1}{2}} = \beta$,
$ \rho_{WW} = (\beta+1)^{2}/(\beta-1)^{2}$, $\kappa_{WW} = 0$ and
$L_{WW}(q^{2}) = B(q^{2})$ [cf.\ \eq{B}].
Also, for the case $j = \gamma$, i.e.\ $M_{j} = 0$, \eq{Lij} reduces to 
\be
L_{i\gamma}(q^{2}) =
\frac{1}{2}\biggl(1 - \frac{M_{i}^{2}}{q^{2}}\biggr)
\ln\biggl(1 - \frac{q^{2}}{M_{i}^{2}}\biggr)
+ \frac{1}{2}\ln 2 \,- 1\,.
\ee

The $ZH$ and $WH$ one--loop contributions to the self--energies
$\hat{\Sigma}_{R,ZZ}$ and $\hat{\Sigma}_{R,WW}$, respectively,
renormalized in the on--shell scheme are then given by
\bea
\label{SigmaRZZZH}
\hat{\Sigma}_{R,ZZ}^{\scriptscriptstyle (ZH)}(q^{2}) &=& 
\frac{1}{\sw^{2}\cw^{2}}
\left.\hat{\Sigma}_{R,ii}^{\scriptscriptstyle (iH)}(q^{2})\right|_{i = Z}\,,\\
\label{SigmaRWWWH}
\hat{\Sigma}_{R,WW}^{\scriptscriptstyle (WH)}(q^{2}) &=& \frac{1}{\sw^{2}}
\left.\hat{\Sigma}_{R,ii}^{\scriptscriptstyle (iH)}(q^{2})\right|_{i = W}\,,
\eea
where
\bea 
\lefteqn{
\hat{\Sigma}_{R,ii}^{\scriptscriptstyle (iH)}(q^{2}) \,\,\,=\,\,\,
\frac{\alpha}{\pi}\Bigl(q^{2} - M_{i}^{2}\Bigr) \times} \nn \\
& &
\Biggl\{ 
\biggl( -\frac{1}{24} - \frac{5M_{i}^{2}}{12q^{2}} + \frac{M_{H}^{2}}{12q^{2}}
- \frac{(M_{i}^{2}-M_{H}^{2})^{2}}{24q^{4}} \biggr)q^{2}
\biggl(\frac{L_{iH}(q^{2})-L_{iH}(M_{i}^{2})}{q^{2}-M_{i}^{2}}\biggr) \nn \\
& &-
\biggl( -\frac{1}{2} + \frac{M_{H}^{2}}{6M_{i}^{2}}
- \frac{M_{H}^{4}}{24M_{i}^{4}} \biggr)M_{i}^{2}
\left.\frac{dL_{iH}(q^{2})}{dq^{2}}\right|_{M_{i}^{2}} 
\nn \\
& &
-\frac{1}{48}\biggl(
1 - \ln \frac{2M_{i}^{2}}{M_{i}^{2}\!+\!M_{H}^{2}}
- \frac{M_{H}^{2}}{M_{i}^{2}}
\biggl[1 - \ln \frac{2M_{i}^{2}}{M_{i}^{2}\!+\!M_{H}^{2}}\biggr] \biggr)
\biggl(1-\frac{M_{H}^{2}}{M_{i}^{2}}\biggr)
\biggl(1-\frac{M_{i}^{2}}{q^{2}}\biggr) \Biggr\} \,.
\eea

Similarly, the $W\gamma$ and $WZ$ one--loop contributions to
the self--energy $\hat{\Sigma}_{R,WW}$ 
renormalized in the on--shell scheme are given by
\bea
\hat{\Sigma}_{R,WW}^{\scriptscriptstyle (W\gamma)}(q^{2})
&=&
\frac{\alpha}{\pi}\Bigl(q^{2} - M_{W}^{2}\Bigr)
\Biggl\{ 
\biggl( \frac{11}{3} - \frac{4M_{W}^{2}}{3q^{2}} 
- \frac{M_{W}^{4}}{3q^{4}} \biggr)q^{2}
\biggl(\frac{L_{W\gamma}(q^{2})-L_{W\gamma}(M_{W}^{2})}{q^{2}-M_{W}^{2}}
\biggr) \nn \\
& &
-2M_{W}^{2}\left.\frac{dL_{W\gamma}(q^{2})}{dq^{2}}\right|_{M_{W}^{2}} 
-\frac{1}{6}\biggl(1 - \ln 2\biggr)
\biggl(1-\frac{M_{W}^{2}}{q^{2}}\biggr) \Biggr\} \,,\\ \nn \\
\hat{\Sigma}_{R,WW}^{\scriptscriptstyle (WZ)}(q^{2}) 
&=&
\frac{\alpha}{\pi}\frac{\cw^{2}}{\sw^{2}}\Bigl(q^{2} - M_{W}^{2}\Bigr)
\Biggl\{
\biggl( \frac{11}{3} - \frac{4M_{W}^{2}}{3q^{2}} + \frac{2M_{Z}^{2}}{3q^{2}}
- \frac{(M_{W}^{2}\!-\!M_{Z}^{2})^{2}}{3q^{4}} \nn \\
& &
+\frac{1}{\cw^{2}}\biggl[-\frac{1}{24} + \frac{19M_{W}^{2}}{12q^{2}}
-\frac{5M_{Z}^{2}}{12q^{2}} 
- \frac{(M_{W}^{2}\!-\!M_{Z}^{2})^{2}}{24q^{4}} \biggr] \biggl)
q^{2}
\biggl(\frac{L_{WZ}(q^{2})-L_{WZ}(M_{W}^{2})}{q^{2}-M_{W}^{2}}\biggr) \nn \\
& &-
\biggl( 2 + \frac{17}{6\cw^{2}} - \frac{2}{3\cw^{4}}
- \frac{1}{24\cw^{6}}\biggl) M_{W}^{2}
\left.\frac{dL_{WZ}(q^{2})}{dq^{2}}\right|_{M_{W}^{2}} 
\nn \\
& &
-\biggl( \frac{1}{6} + \frac{1}{48\cw^{2}}\biggr)
\biggl( \frac{\sw^{2}}{\cw^{2}} \ln\frac{2}{1+\cw^{2}} + \ln \cw^{2}\biggr)
\frac{\sw^{2}}{\cw^{2}}
\biggl(1-\frac{M_{W}^{2}}{q^{2}}\biggr) \Biggr\}\,.
\label{SigmaRWWWgamma}
\eea
Note that in the limit $\sw^{2} = 0$,
$\hat{\Sigma}_{R,WW}^{\scriptscriptstyle (W\gamma)}(q^{2}) = 0$
and, from \eq{SigmaRZZWW},
$\hat{\Sigma}_{R,WW}^{\scriptscriptstyle (WZ)}(q^{2})
= \hat{\Sigma}_{R,ZZ}^{\scriptscriptstyle (WW)}(q^{2})$.

\vspace{20pt}
%\pagebreak

\appendix
\setcounter{equation}{0}
\def\theequation{B.\arabic{equation}}

{\Large {\bf Appendix B}}

\vspace{10pt}

\noi
In this Appendix, we give a brief discussion of the relations between
the renormalization constants involved in the renormalization of
the PT electroweak two--point functions.
In particular, it is described how, as a result of the Ward
identities obeyed by the PT $n$--point functions, 
the three effective charges and the effective weak mixing
angle defined in Sec.~5 in terms of unrenormalized quantities
take exactly the same form when written in terms of the corresponding
renormalized quantities. Thus, they may be related directly to
experimentally measureable quantities, as described in Sec.~7.

Renormalization constants are defined for the electromagnetic coupling
and the gauge boson masses by
\bea\label{constrenorm}
e^{2}     &=& Z_{e}e_{R}^{2}\,,\nn\\
M_{Z}^{2} &=& M_{R,Z}^{2} + \delta M_{Z}^{2}\,, \\
M_{W}^{2} &=& M_{R,W}^{2} + \delta M_{W}^{2}\,,\nn
\eea
and for the gauge fields by
\be\label{fieldrenorm}
\begin{array}{rcl}
\left( \begin{array}{c}
A \\
Z
\end{array} \right) &=&
\left( \begin{array}{cc}
Z_{AA}^{\frac{1}{2}} & Z_{AZ}^{\frac{1}{2}}  \\
Z_{ZA}^{\frac{1}{2}} & Z_{ZZ}^{\frac{1}{2}}
\end{array} \right)
\left( \begin{array}{c}
A_{R} \\
Z_{R}
\end{array} \right)\,,  \\
W^{\pm} &=& Z_{W}^{\frac{1}{2}}W_{R}^{\pm}\,.
\phantom{\displaystyle \frac{M^{M}}{M}}
\end{array}
\ee

The renormalized form $\hat{\Gamma}_{R,NC}$ 
of the matrix $\hat{\Gamma}_{NC}$ \eq{GammaNC} collecting
the PT bare neutral current two--point functions is then given by
\be\label{GammaRNC}
\hat{\Gamma}_{R,NC} = 
\left( \begin{array}{cc}
Z_{AA}^{\frac{1}{2}} & Z_{ZA}^{\frac{1}{2}}  \\
Z_{AZ}^{\frac{1}{2}} & Z_{ZZ}^{\frac{1}{2}}
\end{array} \right)
\hat{\Gamma}_{NC}
\left( \begin{array}{cc}
Z_{AA}^{\frac{1}{2}} & Z_{AZ}^{\frac{1}{2}}  \\
Z_{ZA}^{\frac{1}{2}} & Z_{ZZ}^{\frac{1}{2}}
\end{array} \right)\,.
\ee
Similarly, the renormalized form $\hat{\Gamma}_{R,CC}$ of the PT bare
charged current two--point function is given by
\be\label{GammaRCC}
\hat{\Gamma}_{R,CC}
=
Z_{W}\Bigl(q^{2} - M_{W}^{2} + \hat{\Sigma}_{WW}\Bigr)\,.
\ee

For the particular case of the on--shell scheme considered here,
$\hat{\Gamma}_{R,NC}$ and $\hat{\Gamma}_{R,CC}$ take the form
\bea\label{GammaROS}
\hat{\Gamma}_{R,NC} &=&
\left( \begin{array}{cc}
q^2+\hat{\Sigma}_{R,\gamma\gamma} & \hat{\Sigma}_{R,\gamma Z} \\
\hat{\Sigma}_{R,\gamma Z}         & q^{2} - \bar{s}_{Z} + \hat{\Sigma}_{R,ZZ}
\end{array} \right)\,, \\
\hat{\Gamma}_{R,CC} &=&
q^{2} - \bar{s}_{W} + \hat{\Sigma}_{R,WW}\,,
\eea
where $\bar{s}_{Z}$ and $\bar{s}_{W}$ are the
$Z$ and $W$ complex pole positions, respectively.

It is convenient to introduce as auxilliary quantities
the sine and cosine of the renormalized 
weak mixing angle, together with the corresponding renormalization constants:
\be\label{swrenorm}
\cw^{2} \equiv 1 - \sw^{2} = \frac{M_{W}^{2}}{N_{Z}^{2}} =
\frac{M_{R,W}^{2}+\delta M_{W}^{2}}{M_{R,Z}^{2}+\delta M_{Z}^{2}}
= Z_{\cw}\cwR^{2} \equiv 1 - Z_{\sw}\swR^{2}\,.
\ee
Then, for example, in the on--shell scheme
$\cwR^{2} = M_{R,W}^{2}/M_{R,Z}^{2}$, so that to one--loop order
$Z_{\cw} = 1 + \delta M_{W}^{2}/M_{W}^{2} -\delta M_{Z}^{2}/M_{Z}^{2}$.

The PT one--loop $n$--point functions have been
shown explicitly to obey the same Ward identities as the corresponding
tree level functions \cite{PTfirst,PTrest}. 
The requirement that the renormalized functions
satisfy the same identities as the unrenormalized functions then leads to
relations among the renormalization constants. In particular, 
using these tree--level--like Ward identities, 
it is straightforward to obtain the following relations:
\be\label{Zids}
\begin{array}{rcl}
\left( \begin{array}{cc}
Z_{AA}^{\frac{1}{2}} & Z_{AZ}^{\frac{1}{2}}  \\
Z_{ZA}^{\frac{1}{2}} & Z_{ZZ}^{\frac{1}{2}}
\end{array} \right)
&=&
{\displaystyle \frac{1}{Z_{e}^{\frac{1}{2}}}}
\left( \begin{array}{cc}
1 & \frac{\swR}{\cwR}(Z_{\sw}-1) \\
0 & Z_{\sw}^{\frac{1}{2}}Z_{\cw}^{\frac{1}{2}}\\
\end{array} \right)\,, \\
Z_{W}^{\frac{1}{2}}
&=&
{\displaystyle \frac{1}{Z_{e}^{\frac{1}{2}}}}\, Z_{\sw}^{\frac{1}{2}}\,.
\phantom{\displaystyle \frac{M^{M}}{M}}
\end{array}
\ee

Using the above definitions
of the renormalized quantities $e_{R}$, $\Gamma_{R}$, $\swR$ and $\cwR$,
together with the above relations
among the renormalization constants,
the amplitude for the PT two--point component of the
interaction  between fermions with charges $Q,Q'$ and isospins

\pagebreak

\noindent
$I_{3},I_{3}'$  given in \eq{NC2pt} in terms of bare quantities may be written
\bea
\lefteqn{
\left( \begin{array}{cc}
eQ' & {\displaystyle \frac{e}{\sw\cw}}(I_{3}' - \sw^{2}Q')
\end{array} \right)
\hat{\Gamma}_{NC}^{-1}
\left( \begin{array}{c}
eQ \\ {\displaystyle \frac{e}{\sw\cw}}(I_{3} - \sw^{2}Q)
\end{array} \right) \,\,\,= }\nn \\
& \left( \begin{array}{cc}
e_{R}Q'  & {\displaystyle \frac{e_{R}}{\swR\cwR}}(I_{3}' - \swR^{2}Q')
\end{array} \right)
\hat{\Gamma}_{R,NC}^{-1}
\left( \begin{array}{c}
e_{R}Q \\ {\displaystyle \frac{e_{R}}{\swR\cwR}}(I_{3} - \swR^{2}Q)
\end{array} \right)\,.
\eea
Thus, the amplitude \eq{NC2pt} takes exactly the same form when expressed
in terms of the corresponding renormalized quantities.
As a result, the products
$e^{2}\hat{\Delta}_{\gamma}$
and $(e^{2}/\sw^{2}\cw^{2})\hat{\Delta}_{Z}$
of couplings with the propagator functions
Eqs.~(\ref{PTDeltagamma}) and (\ref{PTDeltaZ}),
and the effective weak mixing angle $\swe^{2}$ \eq{sweff}
defined from the diagonal form of the amplitude on the
r.h.s.\ of \eq{NC2pt},
also take exactly the same form when expressed
in terms of the renormalized quantities.
An exactly similar result holds for the
product $(e^{2}/\sw^{2})\hat{\Delta}_{W}$ in the
PT Dyson--summed two--point component of the 
charge current interaction between fermions given in \eq{CC2pt}.
Alternatively, multiplying out the components of
Eqs.~(\ref{GammaRNC}) and using the relations \eq{Zids},
one can verify the explicit relations
\bea\label{Deltagammarenorm}
q^{2} + \hat{\Sigma}_{R,\gamma\gamma}
&=&
{\displaystyle \frac{1}{Z_{e}}}
\Biggl(q^{2} + \hat{\Sigma}_{\gamma\gamma}\Biggr)\,, \\
\label{DeltaZrenorm}
q^{2} - \bar{s}_{Z} + \hat{\Sigma}_{R,ZZ} 
- \frac{\hat{\Sigma}_{R,\gamma Z}^{2}}
{q^{2} + \hat{\Sigma}_{R,\gamma\gamma}}
&=&
{\displaystyle \frac{Z_{\sw}Z_{\cw}}{Z_{e}}}
\Biggl(q^{2} - M_{Z}^{2} + \hat{\Sigma}_{ZZ} 
- \frac{\hat{\Sigma}_{\gamma Z}^{2}}
{q^{2} + \hat{\Sigma}_{\gamma\gamma}}\Biggr)\,, \\
\label{DeltaWrenorm}
q^{2} - \bar{s}_{W} + \hat{\Sigma}_{R,WW}
&=&
{\displaystyle \frac{Z_{\sw}}{Z_{e}}}
\Biggl(q^{2} - M_{W}^{2} + \hat{\Sigma}_{WW}\Biggr)\,, \\
\label{sweffrenorm}
1  + \frac{\cwR}{\swR}\frac{\hat{\Sigma}_{R,\gamma Z}}
{q^{2}+ \hat{\Sigma}_{R,\gamma\gamma}}
&=&
Z_{\sw}\biggl(1  + \frac{\cw}{\sw}\frac{\hat{\Sigma}_{\gamma Z}}
{q^{2} + \hat{\Sigma}_{\gamma\gamma}} \Biggr)\,,
\eea
among the combinations of self--energies appearing in
$\hat{\Delta}_{\gamma}$, $\hat{\Delta}_{Z}$, $\hat{\Delta}_{W}$
and $\swe^{2}$ in Sec.\ 5
(for clarity, the above expressions have been given in terms
of functions renormalized in the on--shell scheme; they
in fact hold for the components of $\Gamma_{R,NC}$
and $\Gamma_{R,CC}$ in any scheme).

The three effective charges
$\alpha_{\rm eff}$, $\alpha_{Z,{\rm eff}}$ and $\alpha_{W,{\rm eff}}$,
obtained after factoring out the poles
$(q^{2})^{-1}$, $(q^{2} - \bar{s}_{Z})^{-1}$ and
$(q^{2} - \bar{s}_{W})^{-1}$
from the products of bare charges and
diagonal bare propagator functions
$e^{2}\hat{\Delta}_{\gamma}$,
$(e^{2}/\sw^{2}\cw^{2})\hat{\Delta}_{Z}$ and
$(e^{2}/\sw^{2})\hat{\Delta}_{W}$, respectively,
and the effective weak mixing angle $\swe^{2}$,
defined from the diagonal bare neutral current amplitude,
may therefore be expressed entirely in terms of 
the corresponding renormalized quantities,
as in the second equality in each of
Eqs.~(\ref{PTalphaeff})--(\ref{PTalphaWeff}) and (\ref{sweff}).

%\vspace{20pt}
\pagebreak

\appendix
\setcounter{equation}{0}
\def\theequation{C.\arabic{equation}}

{\Large{\bf Appendix C}}

\vspace{10pt}

\noi
In this Appendix, we give the explicit expresssions for the polynomials
$\tilde{F}_{i}(s,x)$, $i = 1,2\ldots 5$, defined in Sec.~7. 
Setting $D\equiv (1-\beta z)(1-z^{2})$, they are as follows:
\bea
\tilde{F}_{1}(s,x) &=& \frac{5}{4D}
\Biggl\{\frac{3}{32}\biggl(16 - (1-\beta z)(71-63z^{2}) \biggr) -3(\beta +z)x
 \nn\\
&& -\frac{21}{16}\biggl(16 - (1-\beta z)(53-45z^{2}) \biggr)x^{2}+
7(\beta +z)x^{3} \nn\\
&& + \frac{63}{32}\biggl(16-(1-\beta z)(43-35z^{2})\biggr)x^{4} \Biggr\}\,,\\
\tilde{F}_{2}(s,x) &=& \frac{315}{128}\Biggl\{3-30x^{2}+35x^{4}\Biggr\}\,,\\
\tilde{F}_{3}(s,x)  &=& \frac{5}{8}\Biggl\{\frac{189}{8}z -21x 
-\frac{945}{4}zx^{2} + 35x^{3} +\frac{2205}{8}zx^{4}\Biggr\}\,,  \\
\tilde{F}_{4}(s,x) &=& \frac{5}{4D}\Biggl\{ -\frac{3}{2}z
+3(1+z^{2})x + 21zx^{2} - 7(1+z^{2})x^{3} - \frac{63}{2}zx^{4}\Biggr\}\,, \\
\tilde{F}_{5}(s,x) &=& \frac{5}{8(1-z^{2})}\Biggl\{ 
\frac{3}{16}(15-70z^{2}+63z^{4}) +3z(5-7z^{2})x 
-\frac{21}{8}(5-42z^{2}+45z^{4})x^{2} \hspace{25pt}
\nn \\
&& - 7z(3-5z^{2})x^{3} + \frac{63}{16}(3-30z^{2}+35z^{4})x^{4}\Biggr\}\,.
\eea
Note that $\tilde{F}_{2}(s,x)$ is proportional to the fourth Legendre
polynomial $P_{4}(x)$.

%\pagebreak

\end{document}